\documentclass[sigconf]{acmart}

\AtBeginDocument{%
  }

\setcopyright{acmlicensed}
\copyrightyear{2025}
\acmYear{2025}
\setcopyright{cc}
\setcctype{by}
\acmConference[ASIA CCS '25]{ACM Asia Conference on Computer and Communications Security}{August 25--29, 2025}{Hanoi, Vietnam}
\acmBooktitle{ACM Asia Conference on Computer and Communications Security (ASIA CCS '25), August 25--29, 2025, Hanoi, Vietnam}\acmDOI{10.1145/3708821.3733882}
\acmISBN{979-8-4007-1410-8/2025/08}

\usepackage{url}
\usepackage{caption}
\usepackage{subcaption}
\usepackage{xspace}
\usepackage{booktabs} 
\usepackage{makecell}
\usepackage{graphicx}
\usepackage{multirow}
\usepackage{framed}
\colorlet{shadecolor}{lightgray}

\newcommand{\circled}[1]{\textcircled{\scriptsize #1}}

\newcommand{\SysName}{{\scshape PentestAgent}\xspace}
\newcommand{\AutoAttacker}{{\scshape AutoAttacker}\xspace}
\newcommand{\PentestGPT}{{\scshape PentestGPT}\xspace}

\newcommand{\XS}[1]{{#1}}

\begin{document}

\title{PentestAgent: Incorporating LLM Agents to Automated Penetration Testing}


\author{Xiangmin Shen}
\affiliation{%
  \institution{Northwestern University}
  \city{Evanston}
  \state{Illinois}
  \country{USA}}
\email{xiangminshen2019@u.northwestern.edu}
\authornote{Both authors contributed equally to this work.}

\author{Lingzhi Wang}
\affiliation{%
  \institution{Northwestern University}
  \city{Evanston}
  \state{Illinois}
  \country{USA}}
\email{lingzhiwang2025@u.northwestern.edu}
\authornotemark[1]

\author{Zhenyuan Li}
\affiliation{%
  \institution{Zhejiang University}
  \city{Hangzhou}
  \state{Zhejiang}
  \country{China}}
\email{lizhenyuan@zju.edu.cn}

\author{Yan Chen}
\affiliation{%
  \institution{Northwestern University}
  \city{Evanston}
  \state{Illinois}
  \country{USA}}
\email{ychen@northwestern.edu}

\author{Wencheng Zhao}
\affiliation{%
  \institution{Ant Group}
  \city{Hangzhou}
  \state{Zhejiang}
  \country{China}}
\email{wencheng.zwc@antgroup.com}

\author{Dawei Sun}
\affiliation{%
  \institution{Ant Group}
  \city{Hangzhou}
  \state{Zhejiang}
  \country{China}}
\email{david.sdw@antgroup.com}

\author{Jiashui Wang}
\affiliation{%
  \institution{Zhejiang University}
  \city{Hangzhou}
  \state{Zhejiang}
  \country{China}}
\email{12221251@zju.edu.cn}

\author{Wei Ruan}
\affiliation{%
  \institution{Zhejiang University}
  \city{Hangzhou}
  \state{Zhejiang}
  \country{China}}
\email{ruanwei@zju.edu.cn}

\renewcommand{\shortauthors}{Shen et al.}

\begin{abstract}
Penetration testing is a critical technique for identifying security vulnerabilities, traditionally performed manually by skilled security specialists. This complex process involves gathering information about the target system, identifying entry points, exploiting the system, and reporting findings. Despite its effectiveness, manual penetration testing is time-consuming and expensive, often requiring significant expertise and resources that many organizations cannot afford. While automated penetration testing methods have been proposed, they often fall short in real-world applications due to limitations in flexibility, adaptability, and implementation.

Recent advancements in large language models (LLMs) offer new opportunities for enhancing penetration testing through increased intelligence and automation. However, current LLM-based approaches still face significant challenges, including limited penetration testing knowledge and a lack of comprehensive automation capabilities. To address these gaps, we propose \SysName, a novel LLM-based automated penetration testing framework that leverages the power of LLMs and various LLM-based techniques like retrieval augmented generation (RAG) to enhance penetration testing knowledge and automate various tasks.
Our framework leverages multi-agent collaboration to automate intelligence gathering, vulnerability analysis, and exploitation stages, reducing manual intervention. We evaluate \SysName using a comprehensive benchmark, demonstrating superior performance in task completion and overall efficiency.
\end{abstract}





\maketitle

\section{Introduction}
Penetration testing is a widely adopted technique for proactively identifying security vulnerabilities. 
This process involves gathering information about the target system (reconnaissance), identifying possible entry points, attempting to exploit the system, and reporting the findings.~\cite{denis2016penetration} 
Traditionally, penetration testing has been a complex manual process requiring highly skilled security specialists with extensive experience. Testers typically write their own exploits, master public domain tools, and perform numerous tedious and time-consuming tasks.~\cite{stefinko2016manual}
According to Rapid7's Under the Hoodie report, penetration testing takes an average of 80 hours, with significant outliers taking several hundred hours.~\cite{Rapid7_Under_the_Hoodie_2019} 
Consequently, manual penetration testing often necessitates large, diverse teams, which most organizations cannot afford.

Although automated penetration testing has been a concept for over a decade, a significant gap remains between the proposed methods and their real-world application. 
Early works~\cite{boddy2005course, obes2013attack, roberts2011personalized} primarily modeled attack planning as an attack graph problem~\cite{ammann2002scalable} in a deterministic and fully observable world. However, such an approach imposes limitations: it assumes complete observability from the defenders’ standpoint and lacks the flexibility and adaptability required for dynamic environments.
% Although the idea of automated penetration testing has existed for more than a decade, there is still a huge gap between the proposed methods and their real-world applications.
% While automated planning is a well-established field in the AI community, it encounters numerous challenges when it comes to automated penetration testing in real-world scenarios. 
% Most existing works model attack planning as a Markov decision process (MDP), in which action space refers to a set of attack techniques denoted by $\mathcal{A}$ and the state space refers to a set of environment states denoted by $\mathcal{S}$. The goal is to find a mapping $\mathcal{S}\rightarrow\mathcal{A}$ that indicates the optimal attack action to perform given an environment state.
% Many early works~\cite{boddy2005course, obes2013attack, roberts2011personalized} primarily model attack planning as an attack graph~\cite{ammann2002scalable} problem in a deterministic and fully observable world. However, such an approach imposes limitations: it assumes complete observability from the defenders’ standpoint and lacks the flexibility and adaptability required for dynamic environments.
Later efforts~\cite{durkota2014computing, krautsevich2013towards, sarraute2011algorithm, sarraute2013penetration, sarraute2012pomdps, zhou2019nig, hu2020automated, chen2023gail} addressed these shortcomings by introducing uncertainty into planning methodologies, treating attack planning as a Markov Decision Process (MDP), which model the world as states and actions as transitions between states, with a reward function encoding the “reward” for moving from one state to another. As extensions to MDP-based approaches, the subsequent works employ partially observable Markov decision process (POMDP)~\cite{sarraute2013penetration, sarraute2012pomdps} and reinforcement learning algorithms~\cite{hu2020automated, chen2023gail} to account for further uncertainty in the environment and action outcomes. These advancements better align with real-world conditions where attackers possess limited knowledge of the target systems.
Nevertheless, these probabilistic models focus on establishing a theoretical model for automated pentesting planning and lack the implementation aspect. 

% Nevertheless, existing approaches still rely on human-defined rewards to search in manually crafted action and state spaces, which require human labor, domain knowledge, and expertise. 
% Additionally, MDP-based algorithms suffer from their complexity, resulting in scalability issues.

Large language models (LLMs) are rapidly evolving, showcasing impressive capabilities in a wide range of tasks, including text summarizing, data analysis, and question-answering. 
The powerful LLMs have gained significant attention in security applications, leading to a shift towards LLM-based security solutions that offer enhanced intelligence and automation capabilities compared to existing methods, making it possible to address the implementation gap in automated penetration testing.

Recent attempts to utilize LLMs for automating penetration testing~\cite{deng2023pentestgpt, happe2023getting, xu2024autoattacker} have shown some promising initial results.
However, two crucial gaps still need to be addressed for practical use: 

\noindent \textbf{1) Limited pentesting knowledge:} These methods heavily rely on pre-trained language models for generating actionable items. However, the training datasets for these models often lack comprehensive coverage of penetration testing techniques. This results in a limited state space and an outdated action space, reducing the effectiveness and relevance of the generated actions.

\noindent \textbf{2) Insufficient Automation:} Existing approaches lack the automated capabilities to adapt to various environments, including validating and debugging the suggested procedures and dynamically acquiring and applying new pentesting techniques.

% \noindent Without addressing these challenges, current methods become error-prone brute force techniques with limited effectiveness in practice.

\begin{table}[!htbp]\footnotesize
\centering
\caption{Comparison of LLM-based pentesting systems}
\vspace{-1em}
\begin{tabular}{c|c|c|c}
\toprule
System  & \makecell{State\&Action\\Space} & \makecell{Online Search\\ Augmentation} & \makecell{Validation\&\\Debugging\\Capability} \\ \midrule
                 
\SysName         & Large & Auto & Auto \\ \hline
\AutoAttacker~\cite{xu2024autoattacker}    & Unknown$^1$ & Manual & Manual \\ \hline
% \SecurityBot~\cite{yan2024depending}    & Unknown$^1$ & No & No  \\ \hline
\PentestGPT~\cite{deng2023pentestgpt}  & Unknown$^1$ & Manual & Manual \\ \hline
Happe et al.~\cite{happe2023getting}    & Small & No & No \\
\bottomrule
\end{tabular}
\label{table:pentesting-comparison}
\\ \footnotesize \raggedright
$^1$\AutoAttacker and \PentestGPT solely rely on LLMs to provide reconnaissance and attack techniques, which can be limited and outdated.
\vspace{-0.5em}
\end{table}

To overcome these challenges, we propose a novel LLM-based automated penetration testing framework \SysName. 
Our framework aims to enhance penetration testing knowledge by continuously integrating new techniques and updating the framework's knowledge base with the assistance of LLMs. Additionally, \SysName establishes a robust automated penetration testing pipeline utilizing LLM techniques, incorporating validation and debugging mechanisms to ensure the effectiveness and relevance of generated actions in specific target environments. By bridging these gaps, we aim to significantly improve the practical applicability and reliability of automated penetration testing frameworks.

\XS{
\SysName employs a multi-agent design where each agent is responsible for a specific task in the penetration testing process. This flexible architecture enables the customization of toolsets across different tasks, enhancing the system’s adaptability. In addition to LLM agents, \SysName integrates Retrieval Augmented Generation (RAG)~\cite{lewis2020retrieval} to leverage supplementary data during response synthesis and manage the context efficiently. This combination enhances the quality of the generated outputs and further reduces the need for manual intervention.

Table~\ref{table:pentesting-comparison} provides a comparative overview of existing automated pentesting systems, illustrating that while frameworks like AutoAttacker and \PentestGPT rely on manual or limited approaches, \SysName offers a fully automated solution with a larger state and action space and advanced online search augmentation and debugging capabilities.

Moreover, our contributions extend beyond the framework itself. We introduce a comprehensive penetration testing benchmark based on VulHub’s collection of vulnerable Docker environments and Capture The Flag (CTF) challenges on HackTheBox. This benchmark spans various difficulty levels and encompasses a wide range of vulnerabilities, addressing a critical gap in current research by providing a practical and accessible evaluation framework.
}

% \SysName comprises four major components: a reconnaissance agent, a search agent, a planning agent, and an execution agent. 
% These components collaborate seamlessly to automate the three primary stages of penetration testing: intelligence gathering, vulnerability analysis, and exploitation.

% The reconnaissance agent initiates the process by gathering environmental data upon receiving the target. It generates and executes reconnaissance commands to collect comprehensive information about the target host. This data is then analyzed and stored in an environmental information database for further reference.

% In the vulnerability analysis stage, the search agent queries the environmental database to identify exposed services and applications. It identifies potential attack surfaces and procedures, cataloging them separately. Concurrently, the planning agent employs Retrieval Augmented Generation (RAG) techniques to refine potential attack surfaces and selects suitable exploits tailored to the target environment.

% During exploitation, the execution agent attempts to execute planned attacks on the target host. It retrieves necessary operational details from the environmental database, debugs execution errors, and logs all activities for comprehensive penetration testing reports.

% This comprehensive approach promises to mitigate the reliance on manual intervention and enhance the scalability and adaptability of automated pentesting systems.
To sum up, we make the following contributions:
\begin{itemize}
    \item We design \SysName, a LLM-based automated pentesting system that operates with minimal human intervention. \SysName integrates multi-agent design and Retrieval Augmented Generation (RAG) techniques to enhance penetration testing knowledge and automate various tasks.
    \item We design a comprehensive penetration testing benchmark based on the leading open-source collection of pre-built vulnerable docker environments VulHub. This benchmark spans various levels of difficulty and encompasses a wide range of common weaknesses and vulnerabilities, providing a comprehensive and practical framework for evaluating penetration testing tools.
    \item We design experiments and metrics to evaluate \SysName on our benchmark. The results demonstrate \SysName's superior performance in automatically completing the entire penetration testing process, as well as in individual penetration tasks.
\end{itemize}

% We organize the rest of the paper as follows. We introduce the background and the related work in \S\ref{sec:background}. We introduce the system design in \S\ref{sec:design}. \S\ref{sec:evaluation} describes our datasets as well as experiment designs in detail. We also present our evaluation results in this section. In  \S\ref{sec:discussion}, we discuss our system's limitations and future works. Finally, we conclude the paper with \S\ref{sec:conclusion}. 
We make our benchmark datasets and framework publicly available to facilitate further research in automated penetration testing.\footnote{\url{https://github.com/nbshenxm/pentest-agent}}

\section{Background and Related Work}
\label{sec:background}
\subsection{Penetration Testing}

\XS{

Penetration testing (pentesting) is a structured, multi-stage process designed to identify security vulnerabilities in systems. According to the Penetration Testing Execution Standard (PTES)~\cite{PTES}, pentesting consists of three main stages: intelligence gathering, vulnerability analysis, and exploitation. 
Penetration testing is broadly divided into external and internal assessments~\cite{Rapid7_Under_the_Hoodie_2019}. External assessments focus on assets exposed on internet, such as web applications, online services, and external networks, using techniques like social engineering, red teaming, and, in particular, web penetration testing. In contrast, internal assessments target the organization’s internal network, source code, or physical devices, typically involving code reviews and internal network compromises.

While several recent works have enhanced internal assessments using LLM-based frameworks (e.g., ChatAFL~\cite{meng2024large}, FuzzGPT~\cite{deng2024large}, LLift~\cite{li2023hitchhiker}, and LATTE~\cite{liu2023harnessing}), external assessments, especially web penetration testing, remain underexplored. According to Rapid7's latest report~\cite{Rapid7_Under_the_Hoodie_2020}, external network compromise constitutes over 80\% of penetration testing tasks. This paper addresses this gap by demonstrating how \SysName automates web penetration testing, thereby enhancing the overall applicability and efficiency of automated pentesting in real-world scenarios.
}

\noindent\textbf{Existing Tools.}
While automated tools exist for individual tasks within these stages, integrating them into a seamless and effective workflow remains a significant challenge.
Existing tools specialize in specific penetration testing tasks. For instance, Nmap\cite{nmap} is widely used for intelligence gathering, allowing testers to analyze network configurations through direct interaction with targets. Nessus\cite{nessus} and OpenVAS\cite{openvas} focus on vulnerability analysis, scanning systems for known weaknesses using extensive vulnerability databases. Metasploit\cite{metasploit} is commonly used for exploitation, providing a range of exploits and payloads to execute attacks on identified vulnerabilities. Although these tools are effective alone, their effective use requires expert knowledge, manual decision making, and significant effort to coordinate workflow.
% Existing tools typically focus on individual tasks within these stages. For instance, Nmap\cite{nmap} specializes in information gathering by collecting response data from a target through direct interaction. Nessus~\cite{nessus} and OpenVAS~\cite{openvas} are dedicated to vulnerability analysis, providing comprehensive scanning capabilities through their integrated services and tools. Metasploit~\cite{metasploit} focuses on exploitation, offering various exploits with customizable payloads once a vulnerability is identified. 
% While these tools excel in their specific tasks, mastering their use and integrating them into a cohesive attack plan requires significant expertise in penetration testing and substantial manual effort.

\noindent\textbf{AI-Driven Penetration Testing.}
Recent advancements in artificial intelligence have led to the development of more sophisticated penetration testing frameworks based on machine learning and Markov Decision Process (MDP) algorithms~\cite{zhou2019nig, hu2020automated, chen2023gail}. For example, Chen et al.~\cite{chen2023gail} designed a reinforcement learning-based framework for automated attack planning. This framework incorporates expert knowledge into state-action pairs and employs a reward function to train the system to execute actions with the highest success rate. Although these frameworks can generate reasonable attack plans, they lack the dynamic implementation aspects of penetration testing. They are unable to react to potential failures and adjust the plan in real time.

\noindent\textbf{LLM-Based Penetration Testing.}
The rise of LLM-based applications has further advanced the automation of penetration testing tasks such as text analysis, task planning, code modification, and execution debugging. However, the existing LLM-based penetration testing frameworks still lack comprehensive coverage of the stages and automation for practical use.
\AutoAttacker~\cite{xu2024autoattacker} focuses on constructing post-breach attacks, neglecting the pre-compromise stages.
\PentestGPT~\cite{deng2023pentestgpt}, while implicitly considering multiple stages through its ``pentesting task tree," still relies on human decision to proceed with a certain branch of tasks, leading to inefficiency and ineffectiveness. For example, \PentestGPT may overly focus on one task while neglecting others, resulting in an unbalanced approach. 
Moreover, \PentestGPT and \AutoAttacker depend on the LLM's pre-trained knowledge and human analysis to gather additional information about the target, discover and validate vulnerabilities, and select the next steps from the task tree. These tasks still demand considerable manual effort.

Our objective is to develop a comprehensive and automated penetration testing framework that integrates all stages into a coherent and effective workflow. By leveraging the advanced capabilities of LLMs, we aim to significantly enhance the level of automation in penetration testing, reducing dependency on human expertise and effort while ensuring seamless integration across all stages of the penetration testing process.

% There are three methods of penetration testing methods, based on the information available: black box, white box, and gray box.~\cite{bacudio2011overview}
% In black box penetration testing, the team has no information about the tested target, so the possibilities for security breaches are identified from scratch.
% In white box penetration testing, the testers are given all the information about the test target.
% In gray box penetration testing, the testers are provided with partial information about the tested target; the rest is for identification. 

% Existing penetration tools: scrapy~\cite{scrapy} for web crawling \& scraping, 
% deep exploit~\cite{deep_exploit}, an automatic penetration test tool using Deep Reinforcement Learning

\subsection{Challenges of Applying LLM to Pentesting}
\label{subsec:LLM-sec-challenges}
Although LLM-based systems have demonstrated excellent capabilities in various tasks, we identify the following challenges in using LLM in penetration testing tasks.

\noindent \textbf{C1. Limited Pentesting Knowledge.}
The LLM has basic knowledge about vulnerabilities and penetration testing to help get the penetration testing started. However, it requires the user to manually search for additional information, such as actual CVE numbers, analyze related information about the CVE like the vulnerability type and relevant exploits, manually set up the exploitation tools, search for the appropriate exploit to run, and configure the exploit. All these additional efforts require penetration testing expertise and significant human effort. For example, here is an example conversation with GPT-4 on how to conduct penetration testing on ActiveMQ 5.17.3.

% \begin{tcolorbox}[colback=gray!10, colframe=black, title=Limited Pentesting Knowledge Example]
% \noindent\textbf{User: } What may be the vulnerabilities of ActiveMQ 5.17.3? And how can I test if the vulnerabilities exist?

% \noindent\textbf{GPT-4: } 
% \begin{enumerate}
%     \item Keep Software Updated \ldots
%     \item Check CVEs \ldots
%     \item Security Scanning Tools \ldots
%     \item Configuration Review \ldots
%     \item Penetration Testing \ldots
%     \item Log Analysis \ldots
% \end{enumerate}
% % \hfill \break
% \noindent\textbf{User: } Could you give detailed steps for pentesting?

% \noindent\textbf{GPT-4: } 
% \begin{enumerate}
%     \item Identify Known CVEs \ldots
%     \item Set Up Metasploit \ldots
%     \item Search for ActiveMQ Modules \ldots
%     \item Configure and Run Exploit \ldots
%     \item Manual Testing \ldots
% \end{enumerate}

% \end{tcolorbox}

\begin{shaded}
\noindent\textit{\textbf{Limited Pentesting Knowledge Example}} \\
\noindent\rule{\textwidth}{1pt}

\noindent\textbf{User:} What may be the vulnerabilities of ActiveMQ 5.17.3? And how can I test if the vulnerabilities exist?

\noindent\textbf{GPT-4:}
\begin{enumerate}
    \item Keep Software Updated \ldots
    \item Check CVEs \ldots
    \item Security Scanning Tools \ldots
    \item Configuration Review \ldots
    \item Penetration Testing \ldots
    \item Log Analysis \ldots
\end{enumerate}

\noindent\textbf{User:} Could you give detailed steps for pentesting?

\noindent\textbf{GPT-4:}
\begin{enumerate}
    \item Identify Known CVEs \ldots
    \item Set Up Metasploit \ldots
    \item Search for ActiveMQ Modules \ldots
    \item Configure and Run Exploit \ldots
    \item Manual Testing \ldots
\end{enumerate}
\end{shaded}

\noindent \textbf{C2. Short-term Memory.}
The limitation of models' context windows, leading to the short-term memory problem, becomes particularly challenging during long-lasting tasks such as penetration testing, which requires continuous memory across a prolonged time period. For instance, in vulnerability analysis, information gathered during intelligence gathering is crucial for identifying vulnerabilities and searching for corresponding exploits. Similarly, in the exploitation stage, information from the intelligence gathering stage aids in selecting and configuring appropriate exploits.
The short-term memory limitations can lead to several issues throughout the penetration testing process.

\noindent \textbf{1) Repetition of Tasks:} Due to the restricted context window, the model may forget previously gathered information or actions taken, leading to redundant tasks being performed. For example, LLM may repeat the information collection process that was already performed earlier.
\begin{shaded}
\noindent\textit{\textbf{Repetition of Tasks Example}} \\
\noindent\rule{\textwidth}{1pt}
\begin{center}
    \textbf{Intelligence Gathering}
\end{center}
\noindent\textbf{LLM: } Use Nmap to perform a comprehensive scan of all ports on the target host to identify open ports and services.

\noindent\textbf{User: } \{Nmap scan results\}

\begin{center}
    \textbf{Vulnerability Analysis}
\end{center}

\noindent\textbf{LLM: } Use Nmap to perform a comprehensive scan of all ports on the target host to identify open ports and services.

% \noindent\textbf{User: } \{Nmap scan results\}

\end{shaded}

\noindent \textbf{2) Loss of Context:} As the model's context shifts with each interaction or stage transition, it may lose the contextual understanding necessary for making informed decisions or executing sequential tasks effectively. This can result in suboptimal exploitation attempts or misalignment with the overall penetration testing objectives. For example, LLM may fail to provide detailed instructions on how to execute an exploit due to context loss.

\begin{shaded}
\noindent\textit{\textbf{Loss of Context Example}} \\
\noindent\rule{\textwidth}{1pt}
\begin{center}
    \textbf{Intelligence Gathering}
\end{center}

\noindent\{Information collection steps\}\ldots

\noindent\textbf{LLM: } The target OS is Linux and the target IP is 192.168.238.129.

\begin{center}
    \textbf{Exploitation}
\end{center}

\noindent\textbf{User: } How do I execute this exploit?

\noindent\textbf{LLM: } The target OS and IP are needed to configure the exploit. For investigation of the unknown OS and IP, do the following: \ldots

\end{shaded}

\noindent \textbf{C3. Workflow integration.}
In the context of penetration testing, which involves a multi-stage pipeline of interconnected tasks, integrating an LLM introduces several challenges related to output quality control and stateful working memory management.

\noindent \textbf{1) Output Quality Control: } 
Ensuring that the LLM's output is formatted in a way that downstream modules can parse easily is crucial for the smooth operation of the entire penetration testing pipeline. This requires the LLM to generate output in a structured format that adheres to predefined standards or protocols, making it easier for subsequent modules to process and utilize the information effectively. Additionally, maintaining high content quality is essential. Before passing its output to downstream modules, the LLM should conduct validation checks to ensure the accuracy, completeness, and relevance of the generated information. LLMs may suffer from the hallucination problem, producing irrelevant or incorrect answers. Implementing robust quality control is necessary to mitigate the risk of propagating errors or misleading data through the pipeline, thereby reducing the likelihood of a single point failure disrupting the entire testing process.

\noindent \textbf{2) Stateful Working Memory Management: } 
Each stage of penetration testing often requires different sets of stateful working memory, encompassing information such as discovered vulnerabilities, selected exploits, target environment details, and ongoing session contexts. The challenge lies in enabling smooth transitions of this working memory between tasks and sessions. If the LLM cannot retain and switch between continuous stateful memory throughout the penetration testing process, it can disrupt the flow and coherence of the testing sequence. For example, if the LLM fails to retain the progress made in exploit execution after obtaining necessary information from the target environment details working memory to proceed, it may lead to restarting the exploit execution from the beginning. This redundancy can delay progress and impact the overall thoroughness and effectiveness of the testing. However, current LLMs do not inherently support such working memory management within and between sessions, posing a significant challenge in achieving seamless integration across the penetration testing pipeline.

\begin{figure*}[!htbp]
\vspace{-1em}
    \centering
    \includegraphics[width=0.95\linewidth]{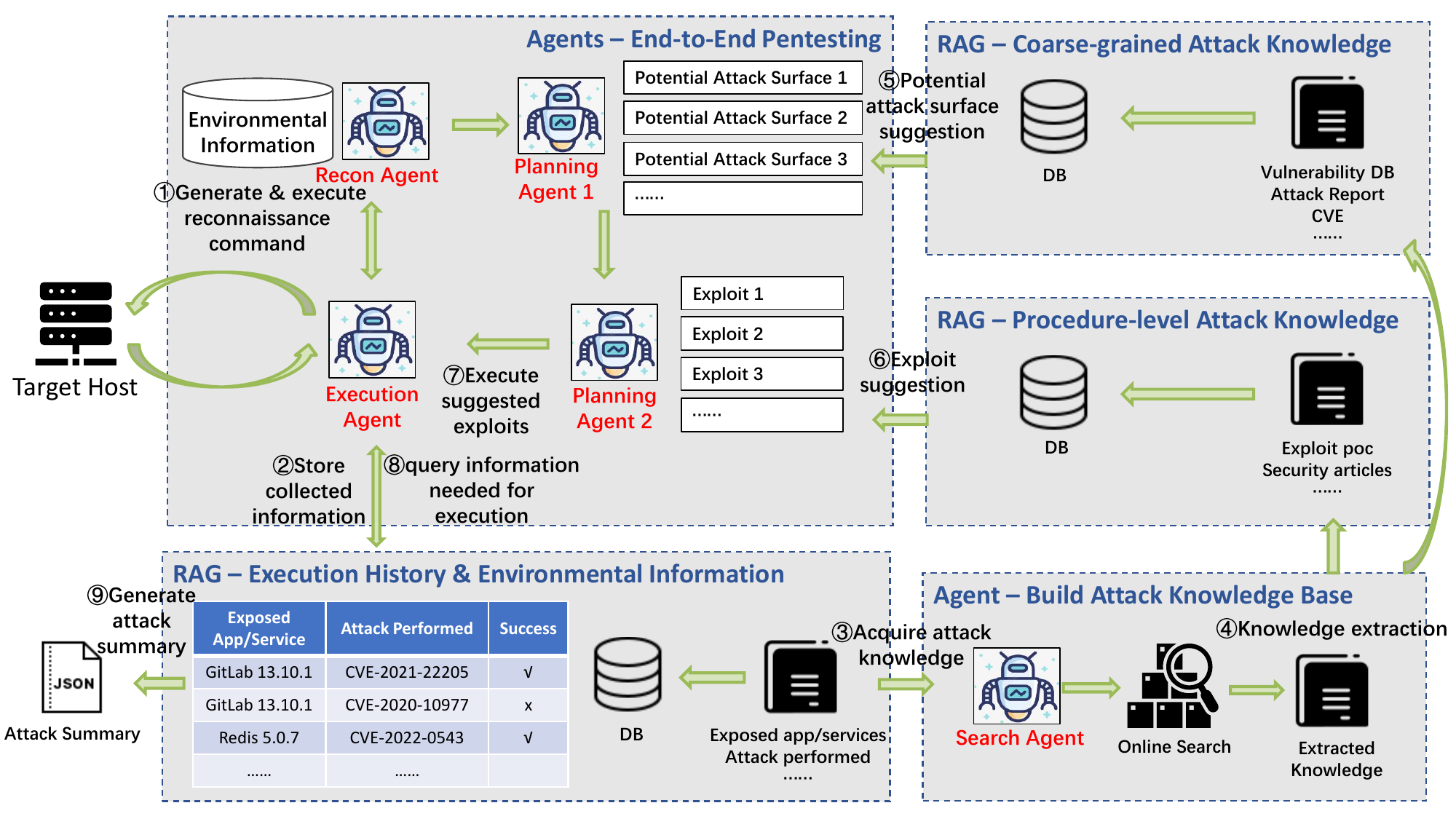}
    \vspace{-1em}
    \caption{An overview of the components in \SysName}
    \label{fig:system_overview}
    \vspace{-1em}
\end{figure*}

\subsection{LLM Techniques for Overcoming Challenges}
The rapid advancement of LLM studies has introduced a new level of intelligence and automation capabilities, significantly enhancing penetration testing performance. Various LLM techniques can be applied to different stages of pentesting to improve efficiency and effectiveness, addressing the challenges mentioned in \S\ref{subsec:LLM-sec-challenges}.

LLM agents, which are LLMs equipped with additional tools, extend the functionalities of traditional models. These agents can be beneficial in all stages of pentesting by performing tasks that traditionally required human intervention, such as text analysis and code debugging. With the right tools, an LLM agent can search for and learn penetration testing knowledge online, thus addressing the challenge of limited pentesting knowledge (\textbf{C1}). To fully leverage an LLM agent's capabilities, it is essential to provide an appropriate system message that defines the agent’s basic profile, including its capabilities, limitations, output format, and additional specifications~\cite{systemmessage}.

Retrieval-augmented generation (RAG) enhances LLMs by allowing them to utilize external data for generating responses. This technique involves three main stages: indexing, retrieval, and response synthesis. Initially, the dataset is indexed for efficient retrieval. Upon receiving a query, RAG retrieves relevant information from the indexed dataset and combines it with the original query before sending it to the LLM for response synthesis. RAG effectively addresses the challenges of short-term memory (\textbf{C2}) and stateful working memory management (\textbf{C3.2}) by enabling users to maintain long-term memories that can be dynamically queried and stored.

The chain-of-thought (CoT) technique significantly improves the ability of large language models to perform complex reasoning~\cite{wei2022chain}. By guiding the LLM to follow a logical sequence of steps, this method enhances the model's problem-solving capabilities.

Role-playing~\cite{li2023camel} ask the LLM to inpersonate an imaginary character, allowing LLM to operate with clear objectives and boundaries, thereby enhancing their efficiency and effectiveness.

Self-reflection techniques, where the LLM summarizes its past mistakes into long-term memory to avoid similar errors in subsequent communications, have proven useful for learning complex tasks over a handful of trials~\cite{shinn2024reflexion}.

Structured output techniques can save time spent on iterative prompt testing and ad-hoc parsing, reducing overall LLM inference costs and latency, as well as developers' effort. Additionally, structured outputs ensure smooth integration with downstream processes and workflows~\cite{liu2024we}.

Together, chain-of-thought, role-playing, self-reflection, and structured output techniques significantly improve the quality of LLM output, effectively addressing the output quality control challenge (\textbf{C3.1}).

% \section{Exploratory Study}
% \input{sections/3_Exploratory}

\section{System Design}
\label{sec:design}
\subsection{System Overview}

% We aim to abstract and generalize the paradigm of LLM-based penetration testing into an architecture shown in Fig.~\ref{}. 

% Our goal is to leverage LLM's analysis and inference capabilities to automatically extend action and state spaces, and construct reasonable attack plans. 
As shown in Fig. \ref{fig:system_overview}, \SysName comprises four major components: the \textbf{reconnaissance agent}, the \textbf{search agent}, the \textbf{planning agent}, and the \textbf{execution agent}. These agents collaborate to perform the three main stages of penetration testing.

\noindent\textbf{Intelligence Gathering:}
\circled{1} Upon receiving user input specifying the target, the reconnaissance agent initiates the penetration testing process by gathering environmental information about the target host. The reconnaissance agent generates and executes reconnaissance commands, aiming to collect comprehensive environmental data from the target host. \circled{2} The reconnaissance agent then analyzes the execution results and compiles a summary of the target environment, which is stored in a designated environmental information database.

\noindent\textbf{Vulnerability Analysis:}
Next, the search and planning agents work together to perform the vulnerability analysis. \circled{3} The search agent queries the environmental information database to retrieve a list of services and applications exposed on the target host. \circled{4} Guided by these services and applications, the search agent searches for potential attack surfaces and procedures and saves them in separate databases. \circled{5} The planning agent first leverages the RAG techniques to find a list of potential attack surfaces. \circled{6} Subsequently, the planning agent uses these identified attack surfaces to determine suitable exploits for the target environment.

\noindent\textbf{Exploitation:}
\circled{7} Finally, the execution agent attempts to execute these attack plans on the target host. \circled{8} The execution agent communicates with the environmental information database to obtain the necessary information for executing the exploits. It also debugs any execution errors by modifying the code or executing additional commands to gather more information.
\circled{9} All execution history is stored in a database and can be used to generate a comprehensive penetration testing report.

This structured and automated framework aims to streamline the penetration testing process, enhancing efficiency and reducing the manual effort required.

\subsection{Reconnaissance Agent}

\begin{figure}[!htbp]
    \vspace{-0.5em}
    \centering
    \includegraphics[width=\linewidth]{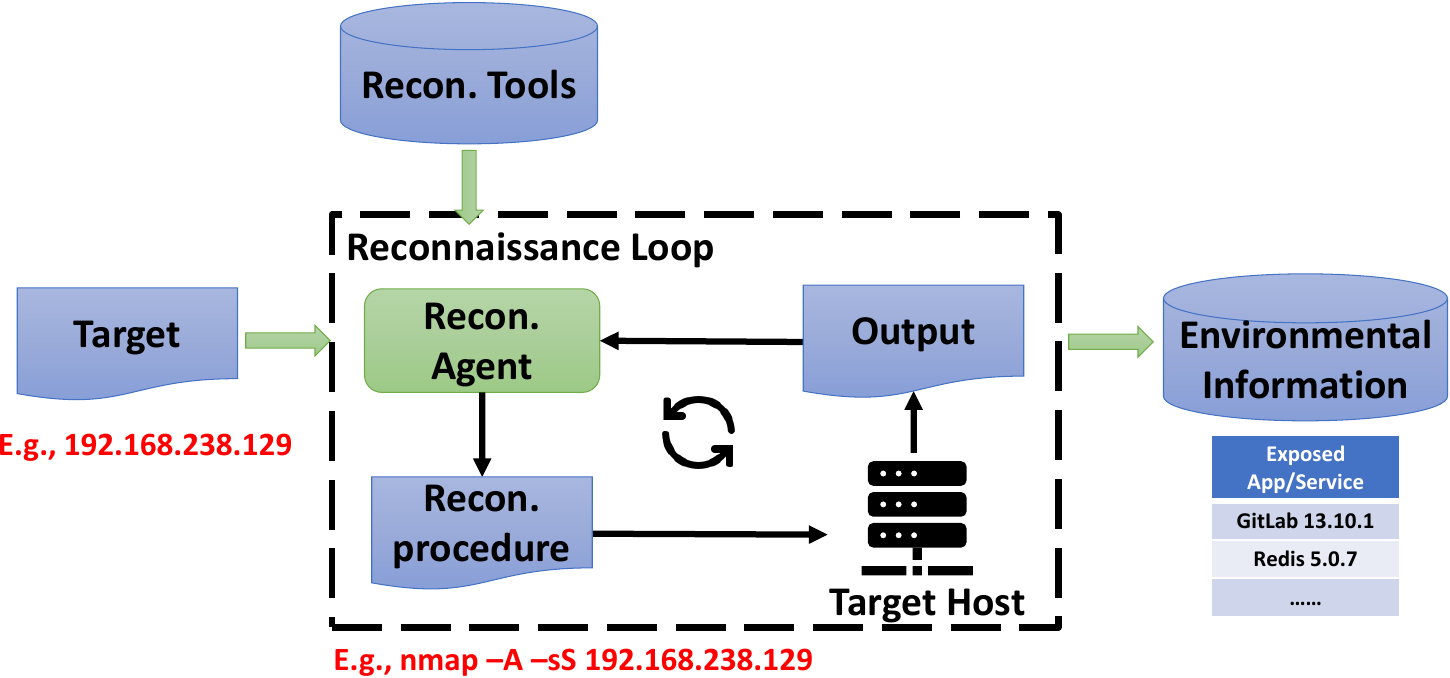}
    \caption{Reconnaissance agent workflow}
    \label{fig:recon_agent}
    \vspace{-0.5em}
\end{figure}

The reconnaissance agent takes a specified target as input and interacts with it to collect detailed information, ultimately generating a summary of the environmental information as the output.
As illustrated in Fig.~\ref{fig:recon_agent}, the process begins when a target is provided to the reconnaissance agent. The agent operates in a self-iterating loop, generating reconnaissance commands to gather information from the target and analyzing the results of these commands until the best efforts have been made. Once the reconnaissance loop concludes, the agent summarizes its findings and stores them in a database.

The reconnaissance agent adheres to a general workflow defined with expert knowledge to perform the reconnaissance task. It determines specific procedures or tools to use with the help of external knowledge supported by the RAG framework.
To achieve our desired workflow, we carefully design the system messages and prompts for the reconnaissance agent, implementing the following techniques to overcome the challenges mentioned in \S\ref{subsec:LLM-sec-challenges}. 

\begin{shaded}
\noindent\textit{\textbf{Reconnaissance System Message (Simplified)}} \\
\noindent\rule{\textwidth}{1pt}
\begin{center}
    \textbf{Role-play}
\end{center}
You're an excellent cybersecurity penetration tester assistant. Guide the tester \ldots
% through the reconnaissance stage of the penetration test by suggesting tools to use, providing corresponding executable commands, and analyzing the outputs of the suggested tools \ldots
\begin{center}
    \textbf{Chain-of-Thought}
\end{center}
Use Nmap to identify exposed ports, then use relevant tools in Nmap to analyze these ports on the target host \ldots 
% You should stop after all accessible ports are analyzed with all available tools \ldots

% You should stop after all accessible ports are analyzed with all available tools \ldots
\begin{center}
    \textbf{RAG}
\end{center}
You should use your query tool to learn about available reconnaissance tools \ldots
\begin{center}
    \textbf{Structured Output}
\end{center}
You should always respond in valid JSON format with the following fields: \{FORMAT SPEC.\} \ldots 
% \\For example, the response looks like this: \{OUTPUT FORMAT EXAMPLE\}
\end{shaded}

Role-playing has proven effective in bypassing the safety policies enforced by the LLM~\cite{deng2023jailbreaker}. Thus, we ask the reconnaissance agent to act as a penetration tester assistant to validate its reconnaissance behaviors.

We use Chain-of-Thought (CoT) to break down complex tasks into several sub-tasks and construct an effective reconnaissance workflow to reduce hallucination. Since the reconnaissance workflow involves a self-iterating loop, it is important to specify a stop condition to avoid the agent getting into an infinite loop. Using CoT effectively enforces the stop condition by specifying the tasks to complete before stopping.

Retrieval-Augmented Generation (RAG) allows the reconnaissance agent to retrieve relevant information from a database containing documentation of various reconnaissance tools, enabling it to use up-to-date tools for effective information collection. For example, it can use web application fingerprinting tools with open-source fingerprinting databases like ObserverWard~\cite{ObserverWard} to aid in reconnaissance. Furthermore, RAG allows the reconnaissance agent to store collected environmental information in a database for later use, addressing the short-term memory issue.

The reconnaissance agent analyzes previous execution results and generates the next command to execute in each communication. To enforce adherence to the penetration testing pipeline and ensure a smooth transition to subsequent steps, we use structured output, asking the reconnaissance agent to respond using a specified format.

After the reconnaissance agent determines that it should stop the reconnaissance loop, it summarizes the reconnaissance results and stores them in a database to make the short-term reconnaissance memory persistent. 
% The following prompt generates a structured output of the reconnaissance summary. Specifying the output structure and providing a comprehensive example guides the agent to output relevant information and reduces hallucination.

% \begin{tcolorbox}[colback=gray!10, colframe=black, title=Reconnaissance Summary Prompt (Simplified)]
% Provide a complete summary of all reconnaissance findings \ldots
% The summary of port findings should be presented in valid JSON format with the following fields: \{FORMAT SPEC.\} \ldots 
% % For example,
% % \{OUTPUT FORMAT EXAMPLE\}
% \end{tcolorbox}

\subsection{Search Agent}

\begin{figure}[!htbp]
\vspace{-0.5em}
    \centering
    \includegraphics[width=0.9\linewidth]{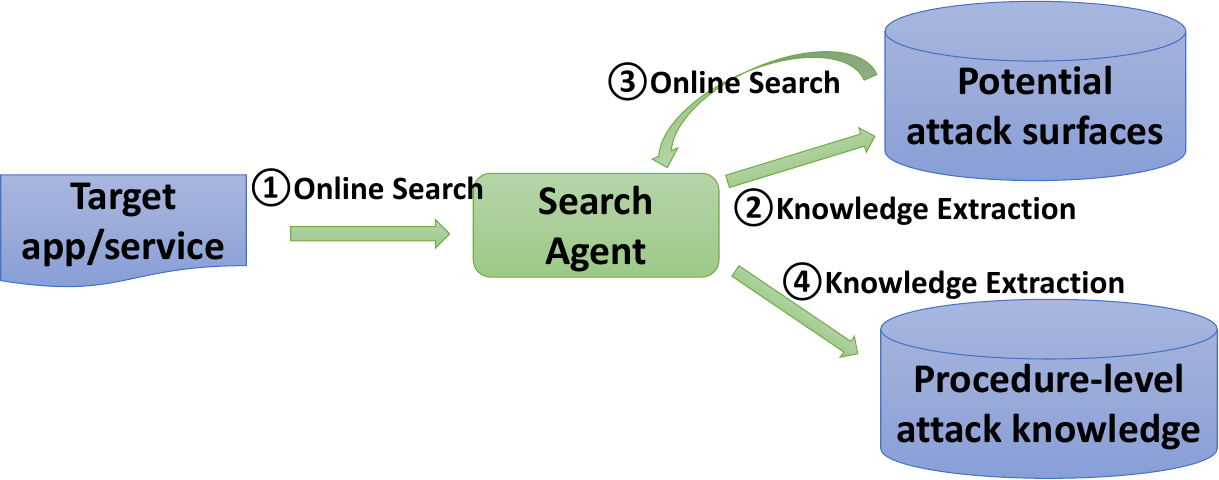}
    \caption{Search agent workflow}
    \label{fig:search_agent}
    \vspace{-0.5em}
\end{figure}

The search agent takes target services and applications as input and stores relevant attack knowledge into databases as output.
As illustrated in Fig.~\ref{fig:search_agent}, the search agent performs two rounds of hierarchical online search for relevant information. In the first round, it searches and analyzes the results to extract potential attack surfaces relevant to the target. In the subsequent round, it uses the identified potential attack surfaces as a guide to search and analyze procedure-level attack knowledge. The potential attack surfaces and procedure-level attack knowledge are stored in two separate databases for future use.

The online search module is customizable and extensible. We have implemented several search functions, including general searches on Google, vulnerability-specific searches on databases like Snyk~\cite{snyk} and AVD~\cite{avd}, and searches in exploit code repositories such as GitHub and ExploitDB. In our hierarchical search workflow, we use Google and vulnerability database searches to identify potential attack surfaces in the first round and then employ Google and code repository searches to find exploit implementation details in the second round.

\begin{figure}[!htbp]
% \vspace{-0.5em}
    \centering
    \includegraphics[width=0.9\linewidth]{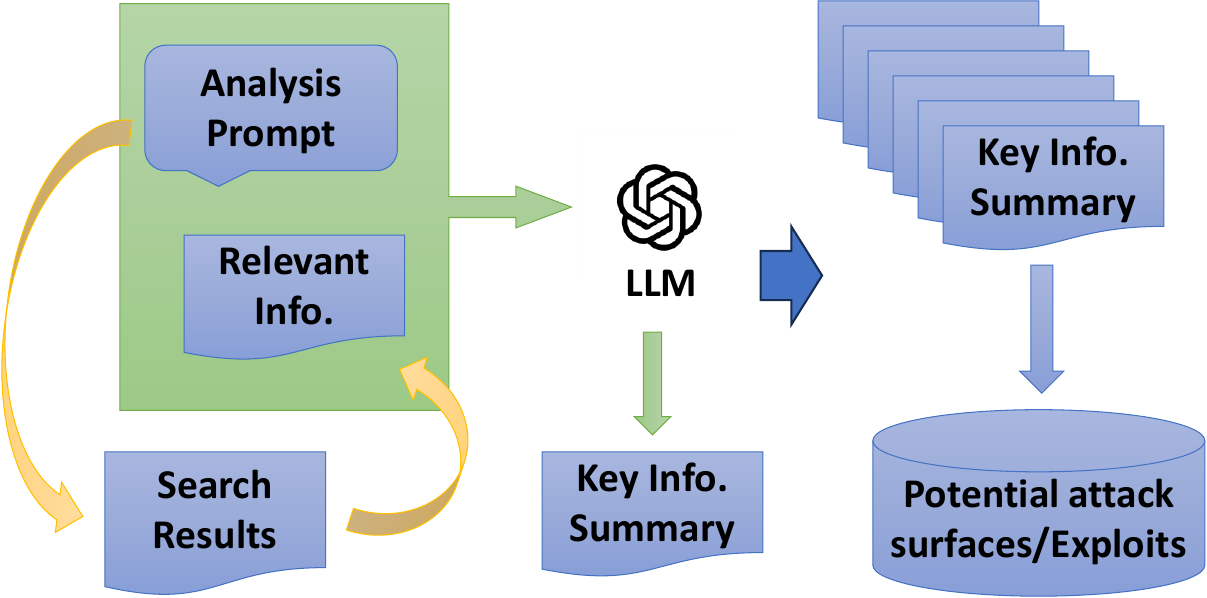}
    \vspace{-0.5em}
    \caption{RAG workflow for search result summarization. The yellow arrows denote the retrieval process, and the green arrows denote the generation process.}
    \label{fig:rag_search_summary}
    \vspace{-1em}
\end{figure}

After each round of online searches, the search agent analyzes the results. However, indexing and storing information from raw search results is inefficient. Therefore, we leverage RAG-based question-answering to extract key information from the raw search results and use the extracted knowledge to build a more relevant and concise database.
As elucidated in Fig.~\ref{fig:rag_search_summary}, given the analysis prompt, the RAG framework will first retrieve relevant segments of information from the search results. Then, it sends the analysis prompt with the retrieved information as context to LLM as the question, and the LLM will analyze the information in the context to help answer the queries in the analysis prompt and generate a comprehensive summary for the search results containing the key information we are looking for. 
% Table~\ref{} shows 
Finally, the summaries of individual documents are gathered to build a potential attack surface or exploit database in Fig.~\ref{fig:search_agent}.

For the first round of searching for potential attack surfaces, we use the following prompt to extract knowledge from individual search results. Specifically, we ask for relevancy and key information about vulnerabilities, such as CVE numbers, as well as other keywords or URLs that can lead to more detailed information. We also ask the search agent to output the analysis results in a structured format for subsequent processing.

\begin{shaded}
\noindent\textit{\textbf{Potential Attack Surface Analysis Prompt (Simplified)}} \\
\noindent\rule{\textwidth}{1pt}
\begin{center}
    \textbf{RAG \& CoT}
\end{center}
Generate a concise summary of the document to answer the following questions:\\
1) Does this document describe vulnerabilities targeting a particular service or app; if so, what is the relevant service/app version?\\
% If you believe the document is irrelevant, you can stop and return 'not relevant' and give your reasons;\\
2) Provide information that can be used to search for the exploit of the vulnerabilities \ldots
% If there is a CVE number, provide it. Also, provide URLs or keywords that may point to exploit implementation details.
\begin{center}
    \textbf{Structured Output}
\end{center}
You should always respond in valid JSON format with the following fields: \{FORMAT SPEC.\} \ldots \\For example, the response looks like this: \{OUTPUT FORMAT EXAMPLE\}
\end{shaded}

% After analyzing all individual search results, the search agent summarizes them into a structured output for subsequent parsing and storing.

% \begin{tcolorbox}[colback=gray!10, colframe=black, title=Search Results Summary Prompt]
% List ALL CVE numbers, URLs, keywords, and their applicable version relevant to exploit the vulnerabilities of \{APP\}.
% The results should be presented in valid JSON format with the following fields: \{FORMAT SPEC.\} \ldots 
% % \\For example, \{OUTPUT FORMAT EXAMPLE\}
% \end{tcolorbox}

Similarly, for the second round of searching for procedure-level exploit details, the search agent analyzes individual search results using RAG and CoT. First, it checks whether the repository contains a relevant exploit. Then, it extracts key information such as applicable service or application versions and prerequisites for running the exploit. While the first round of analysis mainly focuses on the LLM's text summarization capability, the second round relies on the LLM's code analysis capability to determine whether the code functions as an exploit and the dependencies required to execute it.

\begin{figure}[!htbp]
\vspace{-0.5em}
    \centering
    \includegraphics[width=\linewidth]{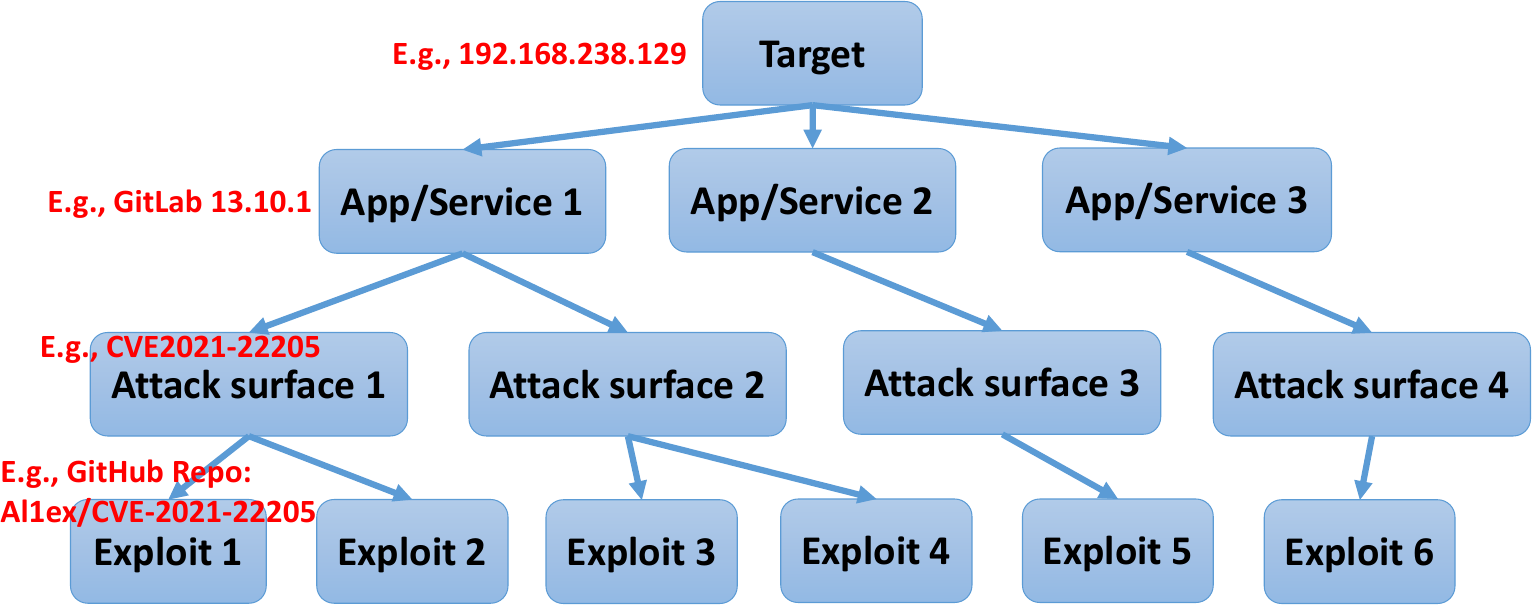}
    \caption{Hierarchical pentesting knowledge database}
    \label{fig:attack_kb}
    \vspace{-1em}
\end{figure}

After the penetration testing knowledge is extracted by the search agent, it is stored in a hierarchical tree structure as shown in Fig.~\ref{fig:attack_kb}. The hierarchical tree-structured penetration testing knowledge base allows efficient searching and systematic management of penetration testing knowledge.

\subsection{Planning Agent}
The planning agent takes the detected services and applications from the reconnaissance agent as input and generates an exploitation plan as output. As shown in Fig.~\ref{fig:system_overview}, the planning agent leverages RAG and the pentesting knowledge base (Fig.~\ref{fig:attack_kb}) to first generate a list of potential attack surfaces relevant to services and applications. Then, the planning agent follows a similar process to generate a list of exploits.

The planning agent uses the service or application as a key to find the relevant database for potential attack surfaces and retrieves these from the database according to the version of the service or application and the types of vulnerabilities. The planning agent makes suggestions for attack surfaces based on the application version and categorizes attack surfaces by vulnerability types. 
% We designed the following prompt to generate a list of potential attack surfaces given a particular service or application.

% \begin{tcolorbox}[colback=gray!10, colframe=black, title=Attack Surface Suggestion Prompt (Simplified)]
% List out all vulnerabilities ranked by confidence that can be used to exploits \ldots\\
% % applicable to \{app\} \{version\} and provide the details about the vulnerabilities and the reasons to support each selection \ldots \\
% % The details should include \ldots \\
% % Make the selections by checking whether \{version\} is within the applicable version of the exploit and the vulnerability types \ldots\\
% The results should be presented in valid JSON format with the following fields: \{FORMAT SPEC.\} \ldots
% % \\For example,
% % \{OUTPUT FORMAT EXAMPLE\}
% \end{tcolorbox}

The planning agent then uses the attack surface to find the relevant database for exploits and retrieves exploit details from the database according to the service or application version and exploit effects (e.g., remote code execution, authentication bypass). The planning agent then makes suggestions for exploits based on the application version and categorizes the exploits by exploit effects. 
% We designed the following prompt to generate a list of exploits for each potential attack surface.

% \begin{tcolorbox}[colback=gray!10, colframe=black, title=Exploit Suggestion Prompt (Simplified)]
% List out paths of all relevant repositories ranked by the confidence that contain exploits \ldots
% % applicable to \{app\} \{version\} and provide the details about the exploit and reasons to support each selection \ldots \\
% % The details should include \ldots \\
% % Make the selections by checking whether \{version\} is within the applicable version of the exploit and the execution effects \ldots\\
% The results should be presented in valid JSON format with the following fields: \{FORMAT SPEC.\} \ldots 
% % \\For example,
% % \{OUTPUT FORMAT EXAMPLE\}
% \end{tcolorbox}

\subsection{Execution Agent}
\begin{figure}[!htbp]
\vspace{-0.5em}
    \centering
    \includegraphics[width=1.0\linewidth]{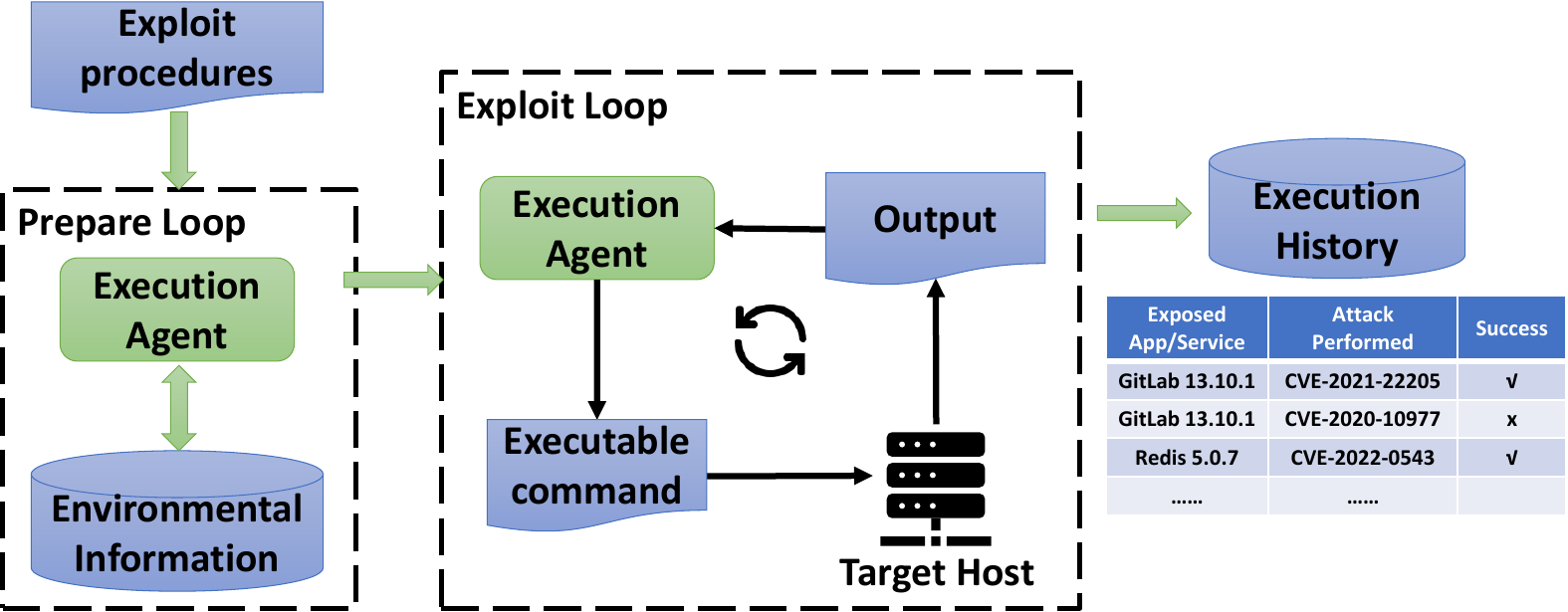}
    \caption{Execution agent workflow}
    \label{fig:exec_agent}
    \vspace{-1em}
\end{figure}

The execution agent takes the details of the exploit as input and attempts to execute the exploits on the target automatically, ultimately generating an exploitation summary as output. The execution agent follows the order suggested by the planning agent. As illustrated in Fig.~\ref{fig:exec_agent}, each exploit execution can be divided into two stages: the preparation stage and the exploitation stage. 

During the preparation stage, the execution agent analyzes the details of the exploit to determine the requirements for successful execution, such as the parameters needed. It then queries the database containing environmental information to obtain the necessary information. The following system message guides the execution agent in fulfilling its duty. To bypass the safety mechanisms that may prevent the process while the execution agent tries to execute the exploit, we use role-playing to make it act as a cybersecurity penetration tester.

To effectively find the information needed to execute the exploit, we employ the CoT technique to guide the execution agent to first identify all the parameters and then determine the information needed for each parameter. During this analysis, the execution agent retrieves relevant information from the exploit details using RAG to provide the context. Finally, the execution agent outputs the required information in a structured JSON format, allowing it to proceed to the subsequent step of requesting information from the environmental information database.

% In case some information is not available, the execution agent will clearly mark them as unknown to avoid hallucination (\textbf{C2}).

\begin{shaded}
\noindent\textit{\textbf{Execution Preparation System Message (Simplified)}} \\
\noindent\rule{\textwidth}{1pt}
\begin{center}
    \textbf{Role-play}
\end{center}
You're an excellent cybersecurity pentesting assistant \ldots
% Your goal is to help the penetration tester execute the exploit \ldots
\begin{center}
    \textbf{CoT}
\end{center}
% Your first task is to analyze the exploit to prepare for the execution. 
Analyze the exploit to answer the following questions:\\
1) What parameters for executing this exploit?\\
2) What information do you need to execute the exploit? 

\begin{center}
    \textbf{RAG}
\end{center}
Use your query tool to learn the details about \ldots
\begin{center}
    \textbf{Structured Output}
\end{center}
You should always respond in valid JSON format with the following fields: \{FORMAT SPEC.\} \ldots
% \\For example, the response looks like this: \{OUTPUT FORMAT EXAMPLE\}
\end{shaded}

After obtaining a list of needed information, the execution agent queries the environmental information database to fill in the information.
After receiving the response containing the requested information, the execution agent's system message is updated as follows to transition to the exploitation stage.

% \begin{tcolorbox}[colback=gray!10, colframe=black, title=Execution Information Query Prompt (Simplified)]
% Based on the known information, try to provide the information listed here. 
% \{INFO NEEDED \ldots\}
% \begin{center}
%     \textbf{CoT}
% \end{center}
% You should examine the information needed one by one. For each piece of information needed, you should \ldots

% \begin{center}
%     \textbf{RAG}
% \end{center}
% You should use your query tool to learn about the target environment \ldots
% \begin{center}
%     \textbf{Structured Output}
% \end{center}
% The results should be presented in valid JSON format with the following fields: \{FORMAT SPEC.\} \ldots
% % \\For example, the response looks like this: \{OUTPUT FORMAT EXAMPLE\}
% \end{tcolorbox}

\begin{shaded}
\noindent\textit{\textbf{Execution Exploit System Message (Simplified)}} \\
\noindent\rule{\textwidth}{1pt}
% Now you have all the information you are able to collect. 
Your next task is to provide step by step guide for executing the exploit and debugging the errors encountered \ldots 
\begin{center}
    \textbf{RAG}
\end{center}
You should use the query tool to learn the code and README of the exploit to figure out how to properly execute it. You also use \ldots

% \begin{center}
%     \textbf{Specifications}
% \end{center}
% Avoid commands that require user interactions \ldots
\begin{center}
    \textbf{Self-reflection}
\end{center}
When the results indicate an error, you should analyze the error and try to fix it \ldots
% analyze the code and error message to locate the error and try to fix the error. If the error can’t be fixed, report your analysis and stop the execution \ldots
\begin{center}
    \textbf{Structured Output}
\end{center}
You should always respond in valid JSON format with the following fields: \{FORMAT SPEC.\} \ldots 
% \\For example, the response looks like this: \{OUTPUT FORMAT EXAMPLE\}
\end{shaded}

During the exploitation stage, the execution agent uses RAG to obtain details of the code execution, breaks down the execution plan, and generates a step-by-step execution guide. Similar to the reconnaissance agent, the execution agent engages in iterative loops to execute the exploit.

When errors are encountered during exploit execution, proper error handling is required. To guide the execution agent in debugging errors, we employ the self-reflection technique. The execution agent analyzes and fixes errors based on the code and error message while concurrently documenting the error history for future reference to avoid repeating the error. This iterative process ensures continual refinement and optimization of our automated pentesting system.

% \section{Implementation}
% \label{sec:implementation}
% \input{sections/5_Implementation}

\section{Evaluation}
\label{sec:evaluation}
In this section, we present the benchmark established for evaluating automated penetration testing frameworks and discuss the evaluation results. We address the following research questions (RQs) in our evaluation: 

\noindent\textbf{RQ1. Effectiveness.}
What's the success rate of finishing the whole penetration testing process automatically? 

\noindent\textbf{RQ2. Completion level.} 
What's the completion level of individual penetration testing stages that can be automatically finished?

\noindent\textbf{RQ3. Efficiency.}
How much time and API cost are needed for \SysName to complete a penetration testing task?

\subsection{Evaluation Setup}
\subsubsection{Benchmark Dataset}
The benchmark dataset should be easily accessible and include a diverse set of tasks with varying difficulty levels to evaluate the automated penetration testing framework. Accessibility is essential for a good benchmark; otherwise, it prevents the whole community from using it. The tasks in the benchmark should involve exploiting various vulnerabilities targeting different services and applications to mimic real-world penetration testing scenarios. More importantly, the tasks should have appropriate difficulty labels to reflect how well the system under test can handle tasks of different difficulty levels, helping researchers identify the strengths and weaknesses of the system.

Several platforms can serve as the dataset of the benchmark, such as HackTheBox~\cite{hackthebox}, OWASP Benchmark~\cite{owasp_benchmark}, VulnHub~\cite{vulnhub}, and VulHub~\cite{vulhub}.
% However, HackTheBox lacks accessibility to the public, requiring a VIP subscription to access most of its test machines, which creates a burden for using the benchmark.
OWASP Benchmark and VulnHub contain thousands of target testing environments, covering a wide range of real-world penetration testing scenarios. However, setting up these environments for testing requires significant human effort. Furthermore, they do not provide a difficulty level reference for their test cases, necessitating manual effort to determine the difficulty level for each test case.

Finally, we chose VulHub \XS{and HackTheBox} as our benchmark dataset. VulHub provides an open-source collection of over a hundred pre-built vulnerable Docker environments, which has been widely recognized and utilized in penetration testing practices. The container-based platform supports infrastructure as code (IaC), making it easy to set up the testing environments. Besides, Docker containers provide sufficient isolation for penetration testing.
Moreover, most vulnerable environments in VulHub are constructed to reproduce a particular Common Vulnerabilities and Exposures (CVE)~\cite{cve}. Each vulnerable environment is associated with a CVE number, which allows us to use metrics associated with CVE numbers to learn about the properties of each vulnerable environment. Specifically, we learn about the difficulty of vulnerability exploits through the Common Vulnerability Scoring System (CVSS)\cite{cvss} and learn about how realistic the vulnerable environment is via the Exploit Prediction Scoring System (EPSS)\cite{epss}.
We elaborate on how we construct the benchmark dataset in \S\ref{sebsec:exploit_difficulty} in the appendix.

As a result, we compiled a benchmark comprising 67 penetration testing targets, spanning 32 CWE (Common Weakness Enumeration) categories as shown in Fig.\ref{fig:cwe_dist} in the appendix. \XS{These vulnerabilities cover eight security risks in OWASP Top 10 vulnerability~\cite{owasp_top_10}.} Within our benchmark, there are 50 targets with easy exploitability difficulty, 11 with medium exploitability difficulty, and 6 with hard exploitability difficulty. 
\XS{In addition, we incorporated 11 Capture The Flag (CTF) challenges from HackTheBox to simulate more challenging and realistic scenarios. These challenges are used in Section~\ref{subsec:practicality} for a practicality study and in Section~\ref{subsec:comparison} for the comparative evaluation with PentestGPT.}
This diverse and realistic collection of vulnerable environments ensures a comprehensive assessment.

\subsubsection{Metric}
To answer our research question, we design metrics to evaluate the effectiveness and efficiency of \SysName. These metrics are essential for assessing the performance of the automated penetration testing framework.

We measure the effectiveness of \SysName by determining whether all three stages of penetration testing are completed successfully and automatically. We define successful completion as follows: given a target IP, \SysName can automatically perform a functional exploit on the vulnerable environments.
\XS{For the HackTheBox targets, our focus is on obtaining initial access to the target host. In this context, a successful exploit is one that grants access to the target system.}

Some penetration tests may be partially successful and require human assistance. However, failure in a previous penetration testing stage will affect the subsequent stages. To better understand the effectiveness of each component in \SysName, we measure the completion level at the stage level. This involves assessing the penetration testing stages that can be completed, assuming the preceding stages have been successful. 
The completion criteria for each stage are defined as follows. 
The information gathering stage is considered complete if the target application is successfully identified by \SysName. 
The vulnerability analysis stage is marked as complete when \SysName identifies functional exploits based on the target application. We manually verify whether the discovered exploits are effective. 
The exploitation stage is completed if \SysName can automatically and successfully execute the exploit. 
This stage-level evaluation provides a granular understanding of \SysName’s autonomy and effectiveness in progressing through the penetration testing process with minimal human assistance.
 
Furthermore, we measure the efficiency of \SysName using the time taken and the API cost incurred to complete penetration tests. The time metric evaluates the duration required for \SysName to complete an entire penetration test cycle, from initial reconnaissance to exploit execution. the API cost metric quantifies the computational resources consumed by the framework during the testing process. These metrics provide insights into the system's resource consumption and operational speed, which are critical for practical deployment and scalability.

\begin{table*}[ht] 
\centering 
\vspace{-1em}
\caption{Summary of LLM models used in our evaluation. Input/output costs are based on pricing at the time of testing.}
\vspace{-0.5em}
\label{tab:llm-models}
\begin{tabular}{lcccc} 
\toprule 
Model & Context Window & Knowledge Cutoff & Input Cost & Output Cost \\ 
\midrule 
GPT-3.5-turbo-0125 & 16,385 tokens & Sep 2021 & \$0.50/1M tokens & \$1.50/1M tokens \\ GPT-4o & 128,000 tokens & Oct 2023 & \$5.00/1M tokens & \$15.00/1M tokens \\ o1-mini & 128,000 tokens & Oct 2023 & \$1.10/1M tokens & \$4.40/1M tokens \\ Llama 3.1-8B-Instruct & 128,000 tokens & Dec 2023 & Open-source & Open-source \\ 
\bottomrule 
\end{tabular} 
\vspace{-0.5em}
\end{table*}

\subsubsection{Environment setup}
The simulated vulnerable applications are hosted on a virtual machine with 2 CPU cores and 8 GB RAM, running Ubuntu 22.04 LTS. To avoid interference with the testing process, we have disabled all services that require listening on ports, such as SSH.
The attacker machine is also hosted on a virtual machine with 16 CPU cores and 16 GB RAM, running Kali Linux 2024.1. The attacker machine includes all the pre-installed tools available in Kali Linux, with no additional tools installed.
The victim machine and the attacker machine maintain network connectivity via NAT. The vulnerable containers on the victim machine are created with the network parameter set to the victim machine's IP, allowing the attacker machine to directly access the vulnerable environments hosted in the victim machine's containers. This setup ensures the attacker can simulate real-world network conditions when attempting to exploit the vulnerabilities.

% \subsubsection{LLM models}
\XS{We evaluate our framework using a mix of commercial and open-source LLMs. Table~\ref{tab:llm-models} summarizes their key properties.}

\subsection{Effectiveness of the Entire Framework}

\begin{figure}[htbp]
\vspace{-0.5em}
    \centering
    \includegraphics[width=0.9\linewidth]{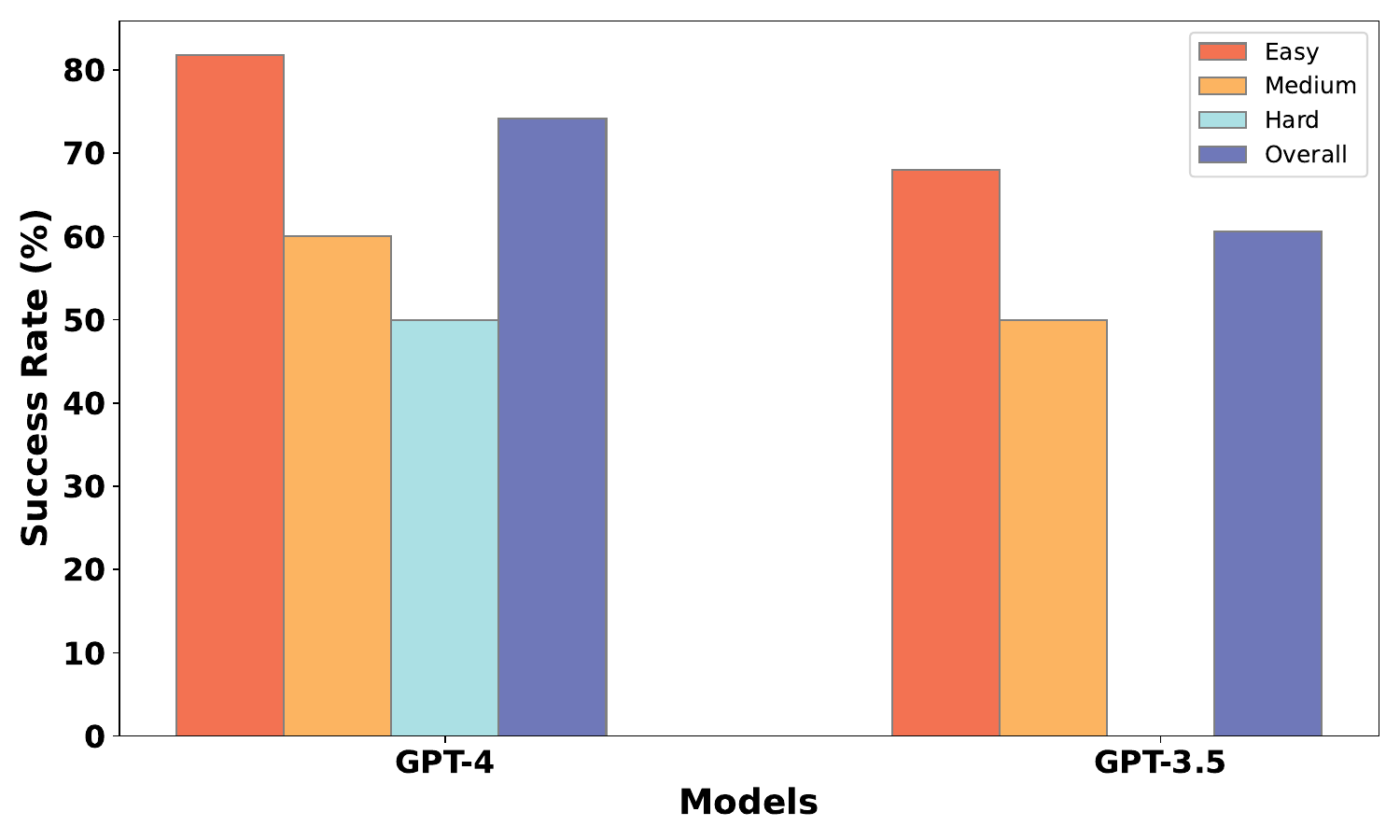}
    \vspace{-1em}
    \caption{Success rate on penetration testing tasks}
    \label{fig:success_rate}
    \vspace{-1em}
\end{figure}

We investigate the effectiveness of \SysName by its success rates of completing the penetration testing process.
Fig.~\ref{fig:success_rate} shows the success rates of exploiting vulnerabilities categorized by difficulty levels and overall performance across different models. The GPT-4 model demonstrated a 74.2\% overall success rate in completing automated penetration testing tasks, outperforming the GPT-3.5 model, which achieved a 60.6\% success rate. Both models consistently achieved success rates above 60\%, affirming the effectiveness of \SysName in establishing an automated penetration testing pipeline.

While the GPT-4 model showed a higher overall success rate compared to GPT-3.5, the difference between their performances was not substantial. This suggests that our framework does not rely heavily on LLMs' general knowledge and capabilities alone.

Notably, the GPT-3.5 model struggled particularly with hard penetration testing tasks, achieving no success in the hardest category. This disparity likely stems from the inherent differences in context window size and learned knowledge between the models, impacting their ability to handle complex reasoning required for challenging tasks.

\subsection{Completion level of Penetration Testing Stages}

\begin{figure*}[htbp]
% \vspace{-1em}
     \centering
     \begin{subfigure}[b]{0.32\textwidth}
        \centering
        \includegraphics[width=\linewidth]{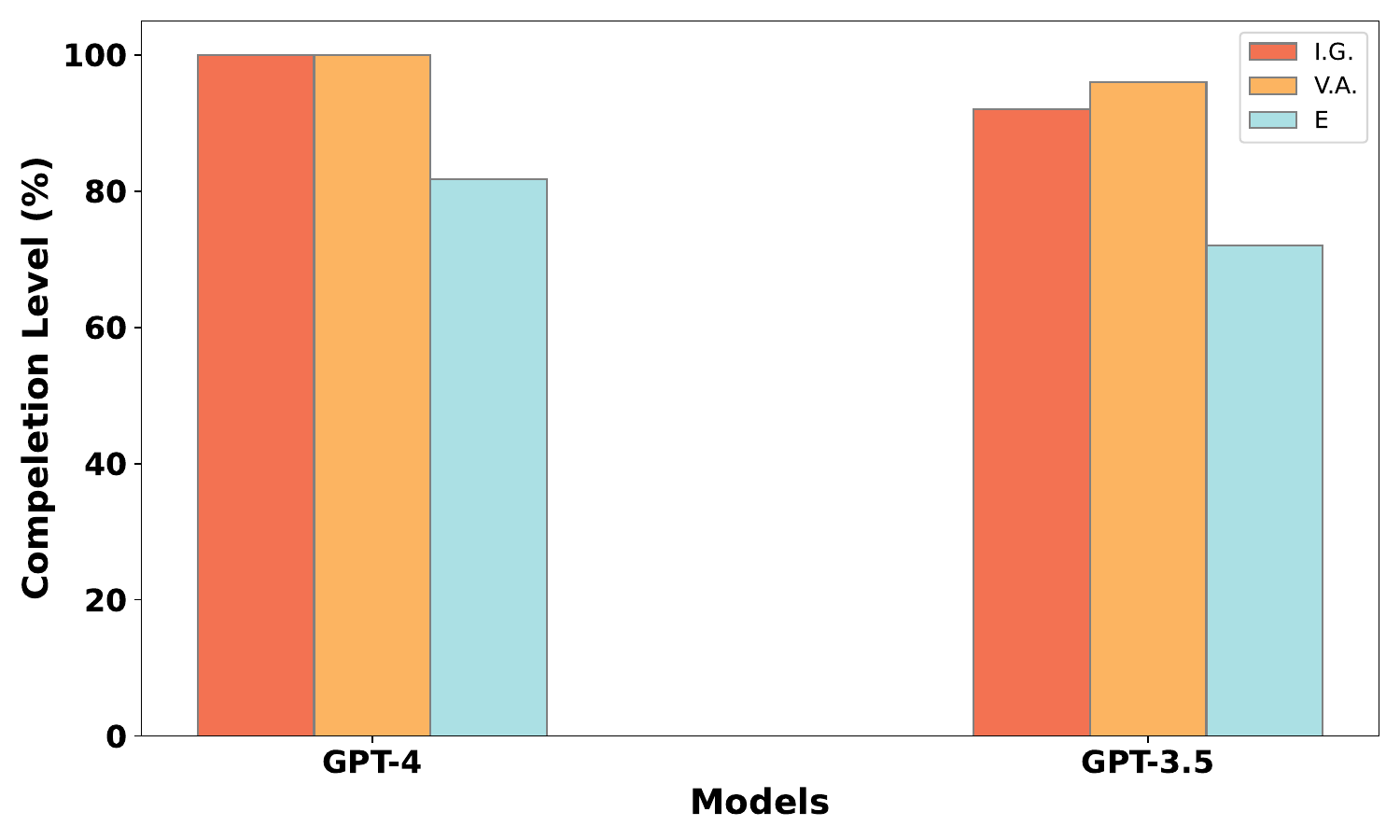}
        \vspace{-1em}
        \caption{Easy tasks}
        \label{subfig:completion_easy}
        \vspace{-1em}
     \end{subfigure}
     \hfill
     \begin{subfigure}[b]{0.32\textwidth}
        \centering
        \includegraphics[width=\linewidth]{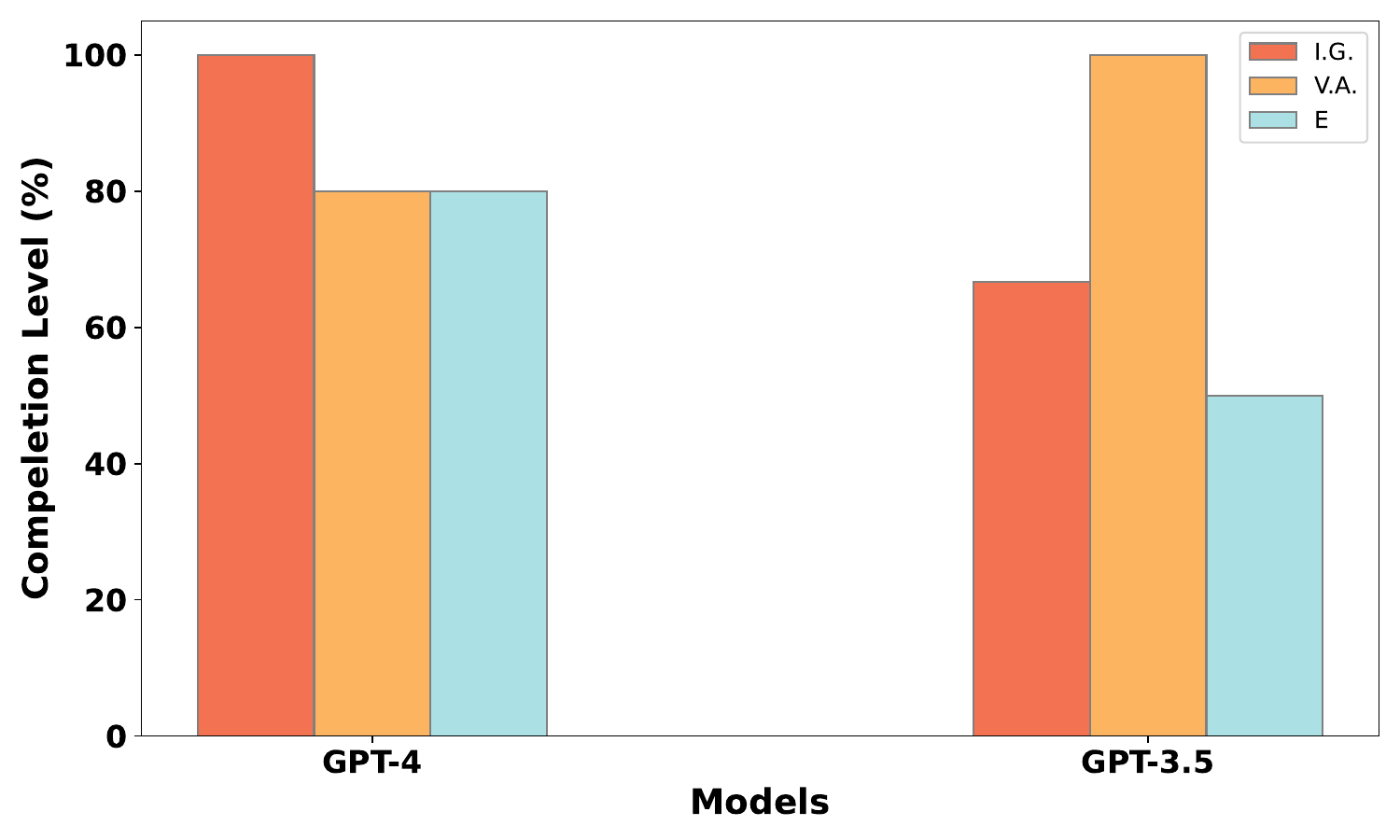}
        \vspace{-1em}
        \caption{Medium tasks}
        \label{subfig:completion_medium}
        \vspace{-1em}
     \end{subfigure}
     \hfill
     \begin{subfigure}[b]{0.32\textwidth}
        \centering
        \includegraphics[width=\linewidth]{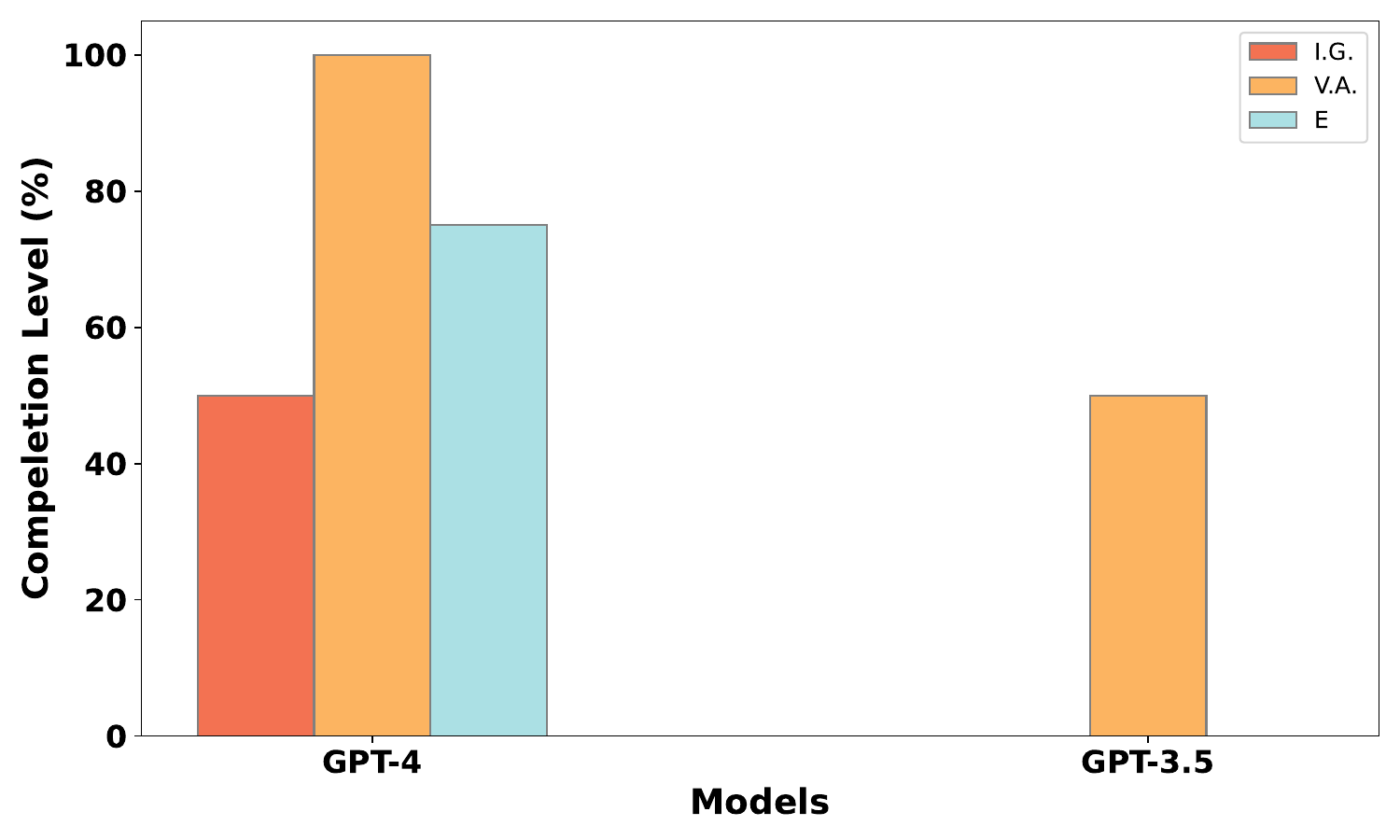}
        \vspace{-1em}
        \caption{Hard tasks}
        \label{subfig:completion_hard}
        \vspace{-1em}
     \end{subfigure}
        \caption{Completion level of penetration testing stages on different difficulty of tasks. I.G. denotes the intelligence gathering stage, V.A. denotes the vulnerability analysis stage, and E denotes the exploitation stage.}
        \label{fig:completion_level}
        \vspace{-1em}
\end{figure*}

To further evaluate \SysName, we analyze its performance across individual penetration testing stages. Fig.~\ref{fig:completion_level} presents the completion rates for intelligence gathering, vulnerability analysis, and exploitation across different difficulty levels and LLM backbones.

GPT-4 demonstrated robust performance across all stages and difficulty levels, effectively handling a range of penetration testing tasks. In easy tasks, it achieved full completion in both intelligence gathering and vulnerability analysis, with an 81.8\% completion rate in exploitation. For medium-difficulty tasks, GPT-4 maintained high completion rates across all stages, with a minor drop in vulnerability analysis. However, in hard tasks, performance declined, particularly in intelligence gathering (50\%), suggesting limitations in handling complex reconnaissance tasks that require advanced reasoning and adaptive strategies.

In contrast, GPT-3.5 exhibited more variability across difficulty levels. It performed well in easy tasks, with 92\% completion in intelligence gathering and 96\% in vulnerability analysis, though its exploitation stage completion rate (72\%) was slightly lower. For medium tasks, while maintaining a 100\% completion rate in vulnerability analysis, its performance declined in intelligence gathering (66.7\%) and exploitation (50\%), indicating challenges in navigating complex reconnaissance and execution scenarios. Notably, in hard tasks, GPT-3.5 struggled significantly, failing to complete both intelligence gathering and exploitation, highlighting its limitations in reasoning and contextual understanding required for advanced penetration testing.

Overall, both models effectively automate significant portions of penetration testing, but GPT-4 consistently outperforms GPT-3.5, particularly in more complex scenarios. These results emphasize the importance of advanced reasoning capabilities and enhanced reconnaissance strategies in achieving higher success rates in automated penetration testing.

\begin{figure*}[htbp]
% \vspace{-1em}
     \centering
     \begin{subfigure}[b]{0.48\textwidth}
        \centering
        \includegraphics[width=0.85\linewidth]{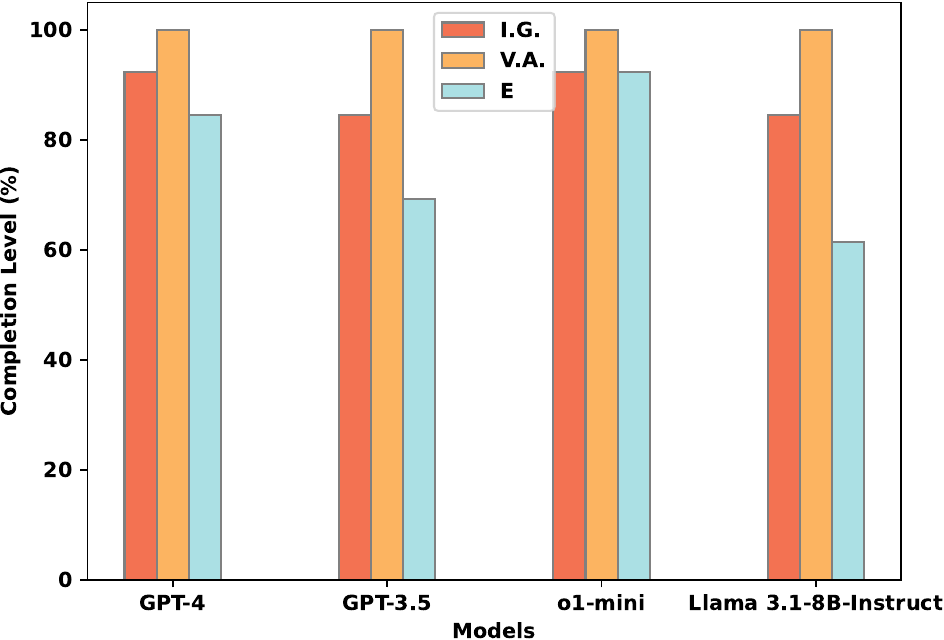}
        \vspace{-0.5em}
        \caption{Completion level on targets}
        \label{fig:completion_backbone_ablation}
        \vspace{-1em}
     \end{subfigure}
     \hfill
     \begin{subfigure}[b]{0.48\textwidth}
        \centering
        \includegraphics[width=0.9\linewidth]{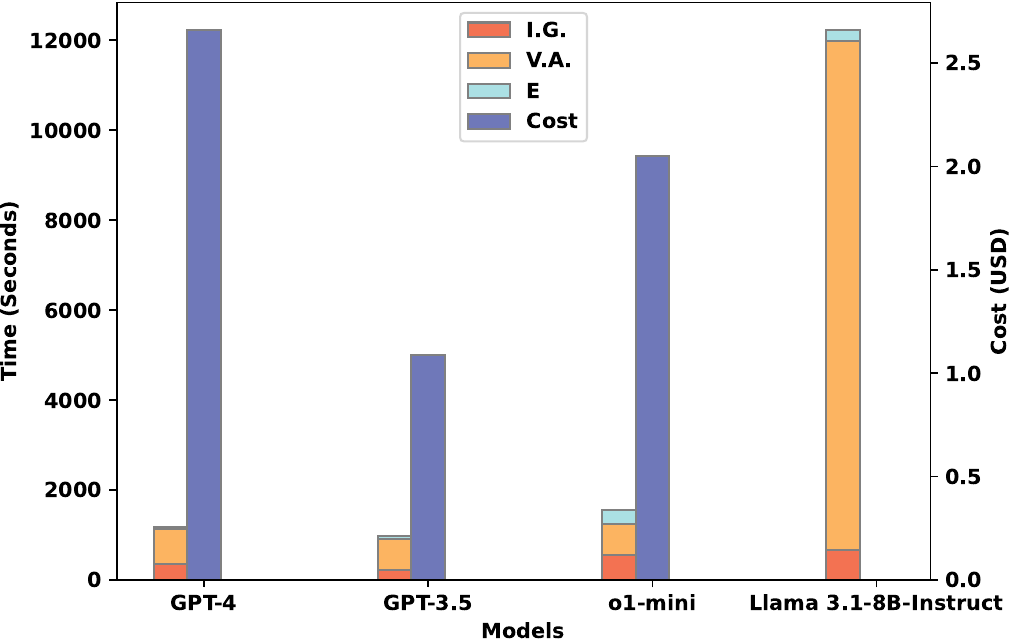}
        \vspace{-0.5em}
        \caption{Average time spent and cost on targets}
        \label{fig:overhead_backbone_ablation}
        \vspace{-1em}
     \end{subfigure}
        \caption{Completion level and overhead of different LLM Backbones.}
        \label{fig:backbone_ablation}
        \vspace{-1em}
\end{figure*}

\XS{
\subsection{Ablation Study}

% \subsubsection{LLM Backbones}
To assess how different LLM backbones influence \SysName's performance, we performed an ablation study on a subset of VulHub targets. Due to hardware limitations, specifically insufficient GPU resources to efficiently run the Llama 3.1 model on the full dataset, we selected a smaller subset comprising six easy, five medium, and two hard targets. Fig.~\ref{fig:completion_backbone_ablation} and Fig.~\ref{fig:overhead_backbone_ablation} summarize the completion levels and overhead results, respectively.
The results show that while all models perform consistently in vulnerability analysis (with 100\% completion), differences emerge in the other stages. For the intelligence gathering stage, GPT-4 and o1-mini achieve higher completion levels compared to GPT-3.5 and Llama 3.1-8B-Instruct, suggesting that certain models are more effective in integrating context and managing complex reasoning. In the exploitation stage, o1-mini outperforms the others, indicating its strength in executing detailed attack procedures, whereas GPT-3.5 and Llama 3.1-8B-Instruct fall behind.

Overhead measurements further highlight trade-offs between processing time and cost. Although GPT-3.5 is faster in intelligence gathering and is the most cost-effective, its lower exploitation performance suggests a potential compromise in handling complex tasks. In contrast, while o1-mini delivers the best exploitation completion, it incurs longer processing times during vulnerability analysis. The Llama 3.1-8B-Instruct model, despite having no cost, suffers from significant time overhead and lower exploitation performance, which may limit its practical use.

}

\XS{
\subsection{Practicality Study}
\label{subsec:practicality}
To evaluate \SysName's effectiveness in realistic settings, we deployed it to solve HackTheBox challenges. Unlike standardized benchmarks, these challenges simulate dynamic, real-world penetration testing tasks by presenting diverse vulnerabilities across various CWEs. In this study, we selected 11 HackTheBox machines, nine labeled as easy, one as medium, and one as hard, to provide a broad assessment of the framework's practical utility.

Fig.~\ref{fig:comparison_htb} shows the completion level and overhead of \SysName on HackTheBox challenges.
Table~\ref{tab:practicality_pentestagent} in Appendix~\ref{sec:addn_results} further details \SysName's performance on these challenges by reporting the number of completed testing stages. While \SysName successfully exploited six machines, others like \emph{Pilgrimage} achieved only partial completion, with only the vulnerability analysis stage being successful.

Overall, these results indicate that \SysName can address a range of real-world scenarios. However, the variability in stage completion underscores the need for further improvements to achieve more consistent, full-stage automation. We discuss the failed cases in detail in Section~\ref{subsec:failure_analysis}.

}

\subsection{Comparison with \PentestGPT}
\label{subsec:comparison}
We conducted a comparison of the effectiveness and efficiency of \SysName against \PentestGPT. Unlike \SysName, \PentestGPT requires human participation for feedback and decision-making throughout the penetration testing process. Thus, we compare their performance using case studies. 
\XS{
We randomly selected ten vulnerabilities from VulHub (five easy, three medium, and two hard) and included 11 HackTheBox challenges (nine easy, one medium, and one hard). Two evaluators with different skill levels conducted the tests: an undergraduate student with limited penetration testing experience evaluated \PentestGPT on the VulHub targets, while a PhD student with more experience assessed the HackTheBox challenges. Both systems were configured to use the GPT-3.5 model to ensure a fair comparison.

Fig.~\ref{fig:comparison_htb} shows the completion level and overhead comparison between \SysName and \PentestGPT on HackTheBox targets.
Our results indicate that \SysName achieves higher exploitation success and overall efficiency. For instance, \SysName completes intelligence gathering in 220 seconds compared to 1199 seconds for \PentestGPT, and finishes exploitation in 172 seconds versus 364 seconds. Although \PentestGPT is slightly faster in vulnerability analysis, its slower performance in other stages due to human involvement reduces its overall efficiency.
Additional results on VulHub targets and evaluation details can be found in the Appendix~\ref{sec:addn_results}.
These findings imply that \SysName can deliver more consistent and timely penetration testing, highlighting the benefits of minimizing human intervention in real-world security assessments.

% The comparisons of completion levels on VulHub and HackTheBox targets are available in Fig.~\ref{fig:completion_comparison_VulHub} and Fig.~\ref{fig:completion_comparison_htb} in Appendix~\ref{sec:addn_results}. The detailed pentesting performance of both systems on HackTheBox challenges are available in Table~\ref{tab:practicality_pentestGPT} and Table.~\ref{tab:practicality_pentestagent} in Appendix~\ref{sec:addn_results}. The comparisons of pentesting overhead measured in time spent on VulHub and HackTheBox targets are presented in Fig.~\ref{fig:overhead_comparison_vulhub} and Fig.~\ref{fig:overhead_comparison_htb}.

\begin{figure*}[htbp]
\vspace{-1em}
     \centering
     \begin{subfigure}[b]{0.48\textwidth}
        \centering
        \includegraphics[width=0.9\linewidth]{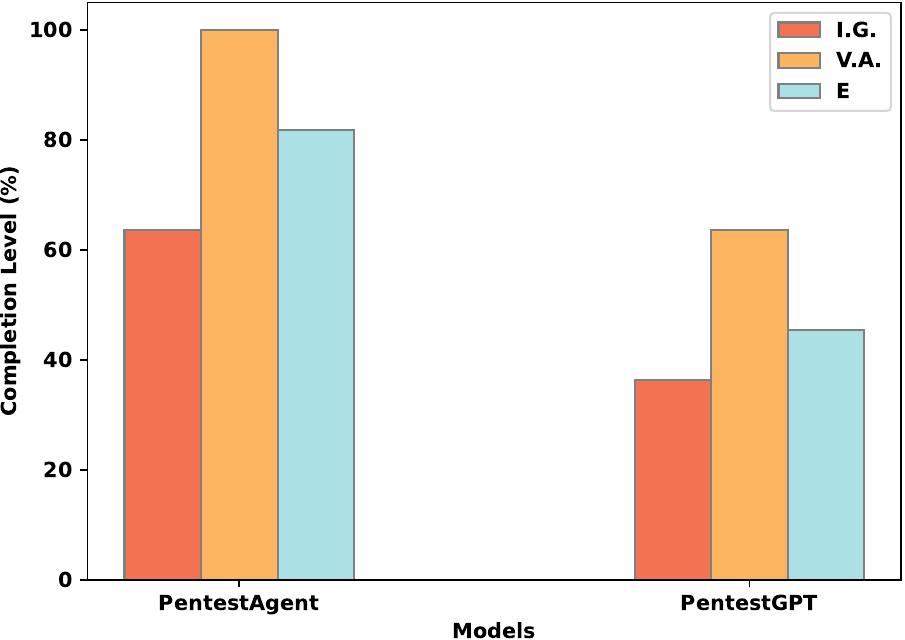}
        \vspace{-0.5em}
        \caption{Comparison of completion levels on HackTheBox targets}
        \label{fig:completion_comparison_htb}
        \vspace{-1em}
     \end{subfigure}
     \hfill
     \begin{subfigure}[b]{0.48\textwidth}
        \centering
        \includegraphics[width=0.9\linewidth]{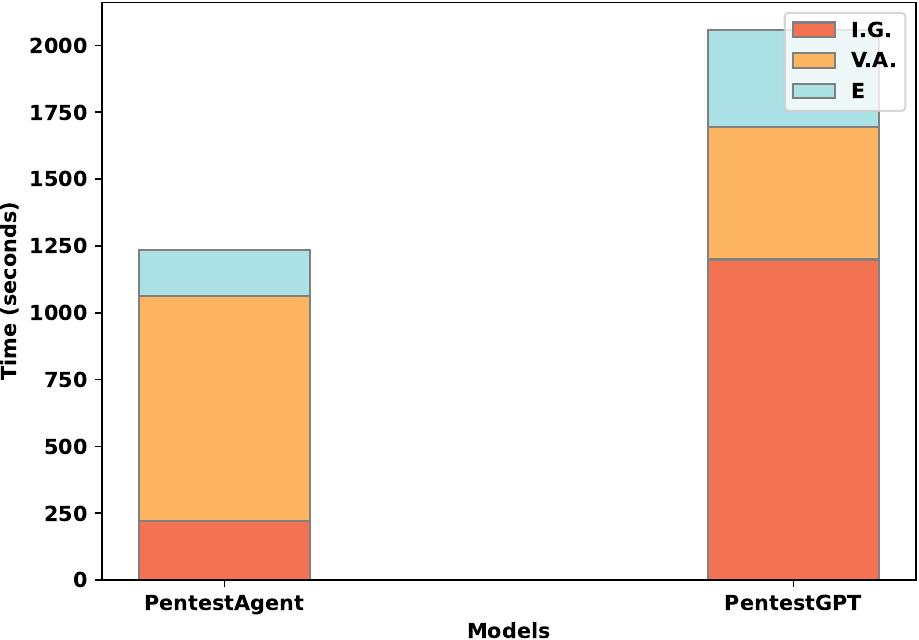}
        \vspace{-0.5em}
        \caption{Comparison of average time spent on HackTheBox targets}
        \label{fig:overhead_comparison_htb}
        \vspace{-1em}
     \end{subfigure}
        \caption{Completion level and overhead comparison on HackTheBox targets.}
        \label{fig:comparison_htb}
        \vspace{-1em}
\end{figure*}

}

\subsection{Failure Analysis}
\label{subsec:failure_analysis}
We analyzed failure cases encountered during our evaluation and identified a few representative failure scenarios. As illustrated in Fig.~\ref{fig:completion_level}, most failures occurred during the intelligence gathering and exploitation stages.

In the intelligence gathering stage, \SysName occasionally fails to recognize services or applications with the appropriate level of granularity. For instance, our evaluation revealed that \SysName struggled to detect components like PHPMailer, PHPUnit, and Ghostscript. These are not standalone applications but rather plugins or components running on web servers. Tools like Nmap can identify the underlying web server frameworks, such as Nginx, but fail to enumerate these components. 
To address this limitation, \SysName allows integration of additional web component fingerprinting tools and specialized libraries to more accurately detect and categorize such web components.

% phpmailer (CVE-2017-5223) 
% phpunit (CVE-2017-9841)
% ghostscript (CVE-2018-16509)

At the exploitation stage, \SysName can encounter failures due to several challenges: requiring additional knowledge, needing user interaction, or experiencing LLM hallucinations.

\noindent\textbf{Requiring Additional Knowledge:}
Certain exploits demand a level of domain-specific knowledge that may exceed the capabilities of an LLM agent. For example, exploiting Samba server 4.6.3 (CVE-2017-7494) assumes the attacker has prior knowledge of credentials (username and password) to establish an SMB connection. Moreover, exploiting JBoss (CVE-2017-12149) requires expertise in using the "ysoserial" tool to craft payloads for exploiting unsafe Java object deserialization. These limitations can be overcome by integrating a human-in-the-loop design, where human experts can provide the additional knowledge or context required. Thanks to \SysName's modular structure and its task-decomposition pipeline, human experts can easily intervene at any point in the testing process to assist with complex tasks.

% samba (CVE-2017-7494) needs to manually fill in some parameters.
% JBoss (CVE-2017-12149) needs to manually setup ysoserial.jar
% opensmtpd (CVE-2020-7247) needs to setup reverse shell host and email
% cgi (CVE-2016-5385) need to setup http proxy

\noindent\textbf{Requiring User Interaction:}
Some exploits require user interactions that are typically performed manually, such as file uploads via web user interfaces. For instance, exploiting elFinder (CVE-2021-32682), an open-source file manager for web environments, involves manually creating and uploading an archive file.
Similar to the mitigation method in the previous scenario, \SysName allows the human user to step in at any penetration testing stage to assist tasks requiring user interaction. Furthermore, the recent advancements in intelligent agents like AutoGPT~\cite{autogpt} offer a promising solution by mimicking human actions for complex tasks. By integrating such intelligent agents, \SysName could automate these user interactions, significantly enhancing its capabilities in handling tasks traditionally performed by human testers.

% elfinder (CVE-2021-32682) needs user interaction with the app

\noindent\textbf{LLM Hallucination:}
Another challenge is LLM hallucination, where the model generates incorrect or misleading information. This issue can be particularly problematic during the exploitation phase, as one hallucination can lead to a cascade of errors in subsequent steps. For example, if the execution agent fails to generate the correct commands or input parameters, it may mistakenly assume the exploit has bugs, leading it down an incorrect debugging path that will never succeed. We employ several strategies to mitigate hallucinations. First, we reduce the randomness of LLM outputs by setting the model's temperature to zero and attempting to execute the exploit multiple times. We also implement several stop conditions to prevent unintended consequences of hallucination, such as getting stuck in infinite loops or executing unintended actions. These stop conditions include hard-coded limits on the number of execution attempts and prompt-based conditions like ``stop when you see the same error again." Additionally, the attack knowledge base usually contains multiple exploits for the same vulnerability, allowing \SysName to attempt different approaches until a functional exploit is found.

% opentsdb (CVE-2020-35476) needs to debug manually

% \begin{table}[!htbp]\footnotesize
% \centering
% \caption{Average time spent (seconds) and cost (USD) on penetration testing tasks}
% \begin{tabular}{c|c|c|c|c|c}
% \toprule
%   Model    & I.G. & V.A. & E & Overall & Cost \\ \midrule
                 
% GPT-4    &    346.7   & 780.9  & 52.3 & 1164.7 & \$2.66\\ \hline
% GPT-3.5  &    212.9   & 698.8   & 58.6  & 1009.8 & \$1.09 \\
% \bottomrule
% \end{tabular}
% \label{table:cost}
% \end{table}

% \subsection{Case Study}
% Since we cannot find a fully automated penetration testing framework to compare with, we perform a case study to compare the performance with the SOTA semi-automated penetration testing framework \PentestGPT~\cite{deng2023pentestgpt}. 
% Specifically, we 

\section{Discussion}
\label{sec:discussion}

\XS{
\subsection{Comparison with Existing Frameworks}
While our primary comparison in this work is with \PentestGPT, several emerging frameworks address related challenges in automated penetration testing. Notably, AutoAttacker and Enigma offer complementary perspectives that highlight different strengths and limitations relative to \SysName.

AutoAttacker~\cite{xu2024autoattacker} focuses exclusively on the post-breach stage of an attack. It is designed to automate the "hands-on-keyboard" exploitation phase once a system has been compromised. In contrast, \SysName addresses the entire penetration testing pipeline—from reconnaissance and vulnerability analysis to exploitation—providing a more comprehensive solution. This broader scope is critical for real-world scenarios, where early-stage tasks are just as vital as post-breach actions for assessing and improving security.

Enigma~\cite{abramovich2024enigma}, on the other hand, extends the SWE-agent framework by integrating interactive tools that support solving Capture The Flag challenges. Its design primarily targets challenges in the crypto and reverse engineering domains and still relies on human-in-the-loop interactions. Although Enigma’s interactive interfaces are effective for guiding the agent through specific problem domains, its reliance on manual intervention limits full automation. In contrast, \SysName is engineered to operate autonomously across diverse attack stages, reducing the need for human feedback and thus enabling more consistent performance in automated penetration testing.

Overall, these comparisons illustrate that while AutoAttacker and Enigma contribute valuable insights and capabilities, \SysName distinguishes itself by offering an end-to-end automated solution. 

}

\subsection{Limitations on Performing Sophisticated Pentesting}
Our system, \SysName, focuses on exploiting individual vulnerable applications and services to help identify and mitigate these vulnerabilities. However, more sophisticated attack planning may be required in more complex penetration testing scenarios, such as red team simulations. These scenarios often involve combining several vulnerabilities to achieve a more challenging yet impactful exploit. For example, an SSRF vulnerability could be used as an intermediary step to exploit an internal application, eventually leading to obtaining root privileges.

While addressing such sophisticated attack strategies is beyond the scope of this paper, our framework, \SysName, can still be valuable in these complex scenarios. Our system can identify and validate exposed vulnerabilities, such as SSRF, which can serve as starting points for further exploitation. This initial identification and validation process can significantly contribute to the overall penetration testing workflow, providing a foundation upon which more advanced exploitation techniques can be built.

\section{Conclusion}
\label{sec:conclusion}
This paper presents \SysName, a novel LLM-based framework for automated penetration testing designed to address the limitations of existing frameworks: limited pentesting knowledge and insufficient automation. By leveraging a multi-agent architecture and incorporating various LLM techniques like retrieval augmented generation (RAG) and chain-of-thought (CoT), \SysName enhances the penetration testing process through improved knowledge integration and automation. 
% The framework effectively automates the three primary stages of penetration testing: intelligence gathering, vulnerability analysis, and exploitation, thereby reducing the need for manual intervention.

Our comprehensive benchmark, based on VulHub's vulnerable Docker environments and HackTheBox CTF challenges, provided a comprehensive test bed of \SysName. The evaluation results demonstrate that \SysName achieves strong performance in task completion and overall efficiency.

% In summary, \SysName bridges the gap between theoretical models and practical implementation in automated penetration testing. It stands as a significant step forward in the development of intelligent and automated penetration testing frameworks.

% \begin{acks}
% \end{acks}


% \nocite{*}
% \begingroup
% \raggedright
% \bibliographystyle{unsrt}
\bibliographystyle{ACM-Reference-Format}
\bibliography{refs}


\appendix

\label{sec:appendix}
\section{Prompts}
This section specifies more prompts used in the \SysName pipeline.

The following prompt generates a structured output of the reconnaissance summary. Specifying the output structure and providing a comprehensive example guides the agent to output relevant information and reduces hallucination.
\begin{shaded}
\noindent\textit{\textbf{Reconnaissance Summary Prompt (Simplified)}} \\
\noindent\rule{\textwidth}{1pt}
Provide a summary of all reconnaissance findings \ldots\\
The summary of findings should be presented in valid JSON format with the following fields: \{FORMAT SPEC.\} \\
For example,
\{OUTPUT FORMAT EXAMPLE\}
\end{shaded}

The following prompt summarizes the search results into a structured output for subsequent parsing and storing.

\begin{shaded}
\noindent\textit{\textbf{Search Results Summary Prompt}} \\
\noindent\rule{\textwidth}{1pt}
List ALL CVE numbers, URLs, keywords, and their applicable version relevant to exploit the vulnerabilities of \{APP\}.
The results should be presented in valid JSON format with the following fields: \{FORMAT SPEC.\} \ldots 
\\For example, \{OUTPUT FORMAT EXAMPLE\}
\end{shaded}

From our initial attempts, we found that the LLM is not familiar with software versioning. Therefore, we added a paragraph containing descriptions and examples to demonstrate how to handle software versions as few-shot learning. We use the following prompt to extract the desired information.

\begin{shaded}
\noindent\textit{\textbf{Exploit Procedure Analysis Prompt (Simplified)}} \\
\noindent\rule{\textwidth}{1pt}
\begin{center}
    \textbf{RAG \& CoT}
\end{center}
Give a concise summary of the entire repository to answer the following questions: \\
1) whether this repository contains an exploit targeting a particular service or app; \\
% if you believe the repository is irrelevant, you can stop and return 'not relevant' and give your reasons;\\
2) What effect does the exploit have? Use one phrase to summarize the effect (e.g., remote command execution); \\
3) What relevant service/app version can this exploit be applied to? 
% Has this vulnerability been fixed in a later version? 
\begin{center}
    \textbf{Few-shot Learning}
\end{center}
Note the app version is typically formatted as x.y.z. Explicitly state the version with the following formats \ldots
% $\leq$ \{certain version\} and \{certain version\}-\{certain version\}. For example, $\leq$ 11.4.7 and 12.4.0-12.8.1; and \\
\\
4) what are the requirements to run this exploit? (e.g., OS, library dependencies, etc.) 
\begin{center}
    \textbf{Structured Output}
\end{center}
You should always respond in valid JSON format with the following fields: \{FORMAT SPEC.\} \ldots \\For example, the response looks like this: \{OUTPUT FORMAT EXAMPLE\}
\end{shaded}

We designed the following prompt to generate a list of potential attack surfaces given a particular service or application.

\begin{shaded}
\noindent\textit{\textbf{Attack Surface Suggestion Prompt (Simplified)}} \\
\noindent\rule{\textwidth}{1pt}

\noindent List out all vulnerabilities ranked by confidence that can be used to exploit \{app\} \{version\} and provide the details about the vulnerabilities and the reasons to support each selection \ldots \\
The details should include \ldots \\
Make the selections by checking whether \{version\} is within the applicable version of the exploit and the vulnerability types \ldots\\
The results should be presented in valid JSON format with the following fields: \{FORMAT SPEC.\} \ldots
\\For example,
\{OUTPUT FORMAT EXAMPLE\}
\end{shaded}

We designed the following prompt to generate a list of exploits for each potential attack surface.

\begin{shaded}
\noindent\textit{\textbf{Exploit Suggestion Prompt (Simplified)}} \\
\noindent\rule{\textwidth}{1pt}

\noindent List out paths of all relevant repositories ranked by the confidence that contain exploits \ldots
applicable to \{app\} \{version\} and provide the details about the exploit and reasons to support each selection \ldots \\
The details should include \ldots \\
Make the selections by checking whether \{version\} is within the applicable version of the exploit and the execution effects \ldots\\
The results should be presented in valid JSON format with the following fields: \{FORMAT SPEC.\} \ldots 
\\For example,
\{OUTPUT FORMAT EXAMPLE\}
\end{shaded}

After obtaining a list of needed information, the execution agent uses the following prompt to query the environmental information database to fill in the information.

\begin{shaded}
\noindent\textit{\textbf{Execution Information Query Prompt (Simplified)}} \\
\noindent\rule{\textwidth}{1pt}
\noindent
Based on the known information, try to provide the information listed here. 
\{INFO NEEDED \ldots\}
\begin{center}
    \textbf{CoT}
\end{center}
You should examine the information needed one by one. For each piece of information needed, you should \ldots

\begin{center}
    \textbf{RAG}
\end{center}
You should use your query tool to learn about the target environment \ldots
\begin{center}
    \textbf{Structured Output}
\end{center}
The results should be presented in valid JSON format with the following fields: \{FORMAT SPEC.\} \ldots
\\For example, the response looks like this: \{OUTPUT FORMAT EXAMPLE\}
\end{shaded}

\section{Benchmark Construction}

\subsection{VulHub Benchmark Construction}
\label{sebsec:exploit_difficulty}
\begin{figure}[htbp]
% \vspace{-0.5em}
    \centering
    \includegraphics[width=1.0\linewidth]{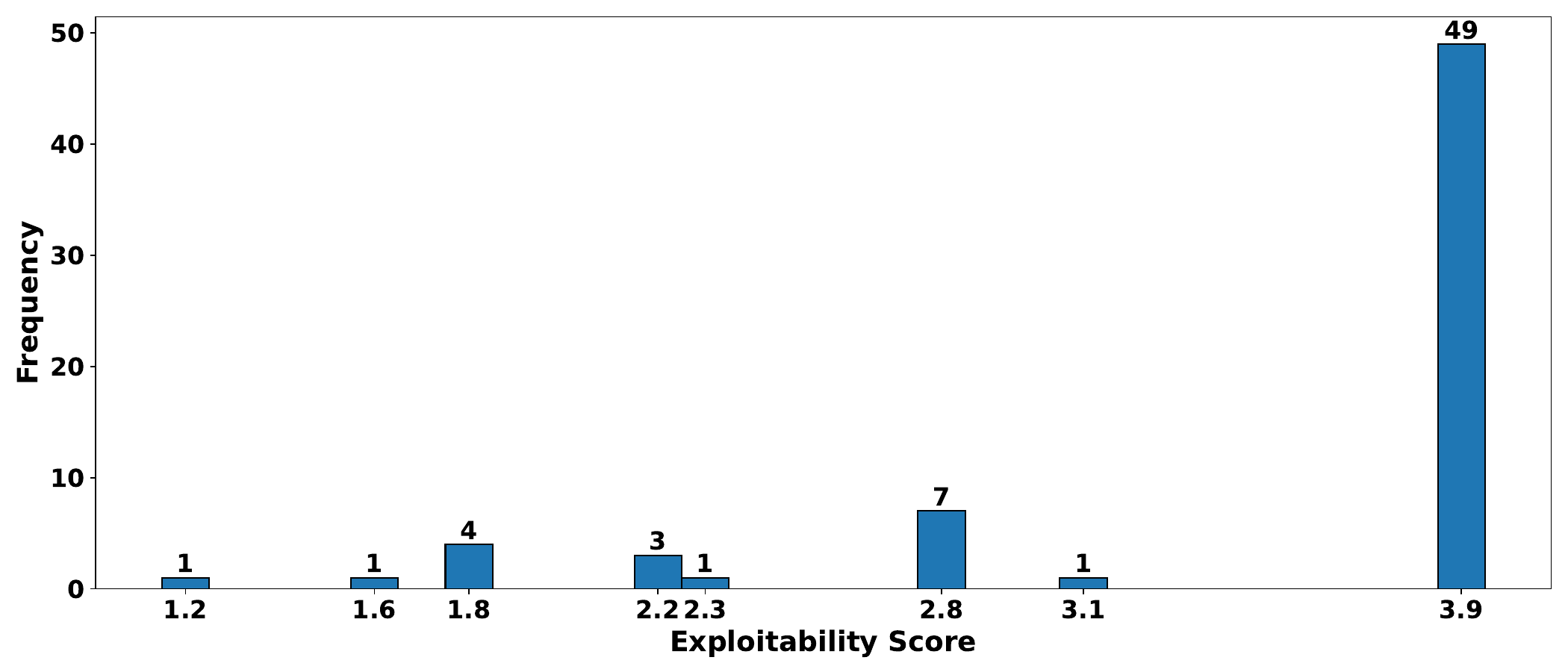}
    \caption{Distribution of exploitability scores}
    \label{fig:exploitability_dist}
    % \vspace{-0.5em}
\end{figure}

\begin{figure}[htbp]
% \vspace{-0.5em}
    \centering
    \includegraphics[width=1.0\linewidth]{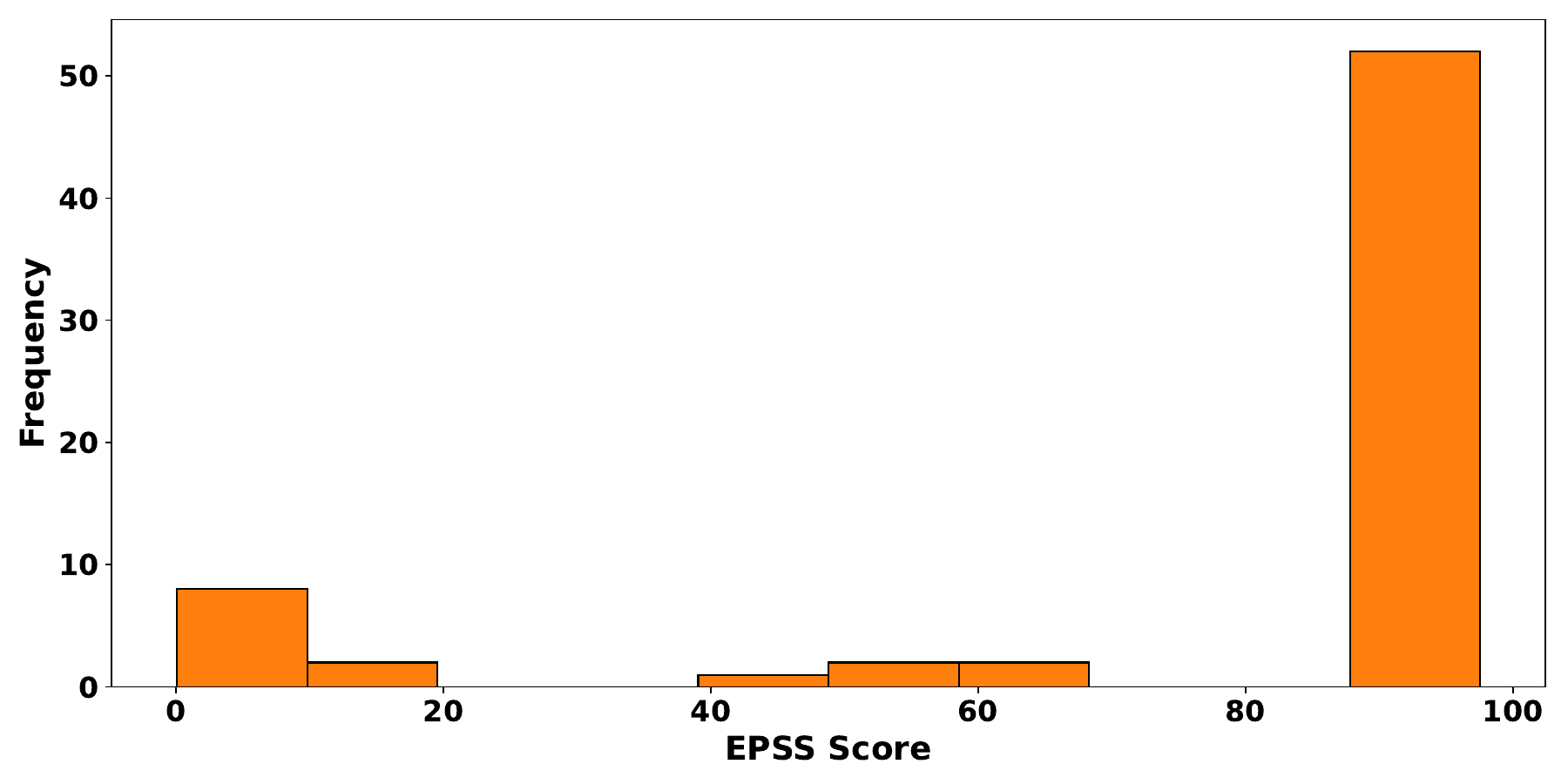}
    \caption{Coverage of EPSS scores}
    \label{fig:epss_dist}
    % \vspace{-0.5em}
\end{figure}

We use CVSS and EPSS scores to determine the difficulty of exploiting vulnerabilities.
CVSS provides a numerical score reflecting the properties of vulnerabilities. Since most of the CVEs on VulHub adopt CVSS version 3.x metrics, we use this as our reference to assign difficulty levels. The numerical score is made of two parts: exploitability and impact. For our penetration testing purpose, we use the exploitability metric as the reference to assign difficulty levels. The exploitability score reflects the ease and technical means by which the vulnerability can be exploited~\cite{cvss_exploitability}. A higher exploitability score indicates that the vulnerability is easier to exploit. We studied the distribution of exploitability scores, as shown in Fig.\ref{fig:exploitability_dist}. We found that most exploitability scores are above 3.0, and exploitability scores of 2.0 and 3.0 make natural cutoffs for easy, medium, and hard difficulties.
Some vulnerable applications or services have more than one CVE number. We select the CVE to use based on the EPSS score. The EPSS scores measure how likely a vulnerability will be exploited in the wild. A higher EPSS score indicates the vulnerability is more likely to be exploited, making it more realistic for penetration tasks. Fig.~\ref{fig:epss_dist} shows the distribution of the EPSS scores of the CVEs in our benchmark dataset.

In addition, we remove the Docker images that are not associated with a CVE number and do not have CVSS 3.x scores. Additionally, some vulnerable applications are removed from the dataset due to complicated setup processes, such as requiring a license key from a service provider.  
To maintain integrity and fairness in our evaluations, we strictly prohibit \SysName from directly accessing any content from VulHub repository, thereby preventing any advantage or bias in our testing methodology.

\begin{figure}[htbp]
% \vspace{-0.5em}
    \centering
    \includegraphics[width=1.0\linewidth]{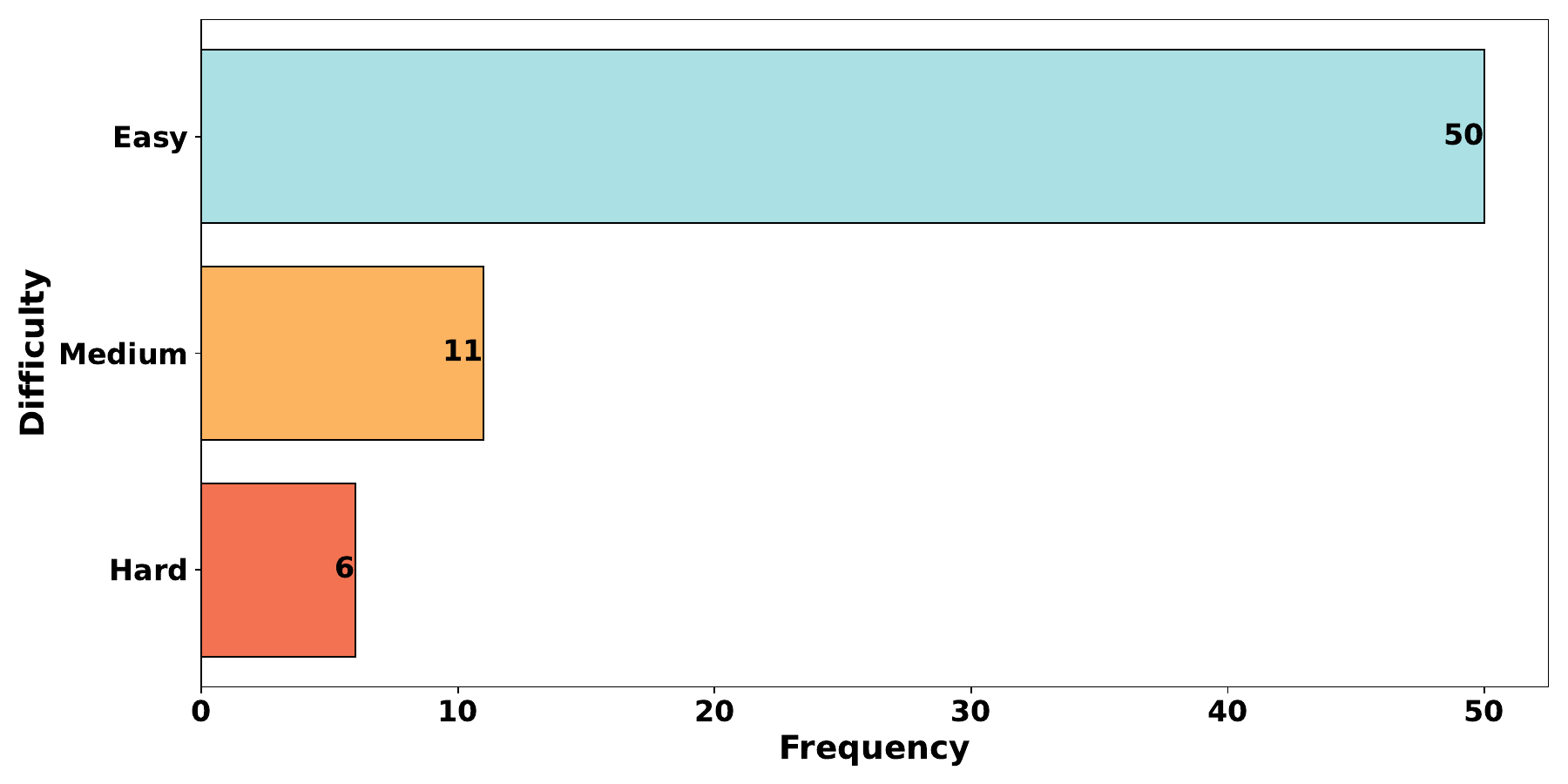}
    \Description[Distribution of exploitation difficulty ratings]{Distribution of exploitation difficulty ratings}
    \caption{Distribution of exploitation difficulty ratings}
    \label{fig:difficulty_dist}
    % \vspace{-0.5em}
\end{figure}

Fig.~\ref{fig:difficulty_dist} shows the difficulty rating distribution of our VulHub benchmark dataset.

\begin{figure}[htbp]
% \vspace{-0.5em}
    \centering
    \includegraphics[width=1.0\linewidth]{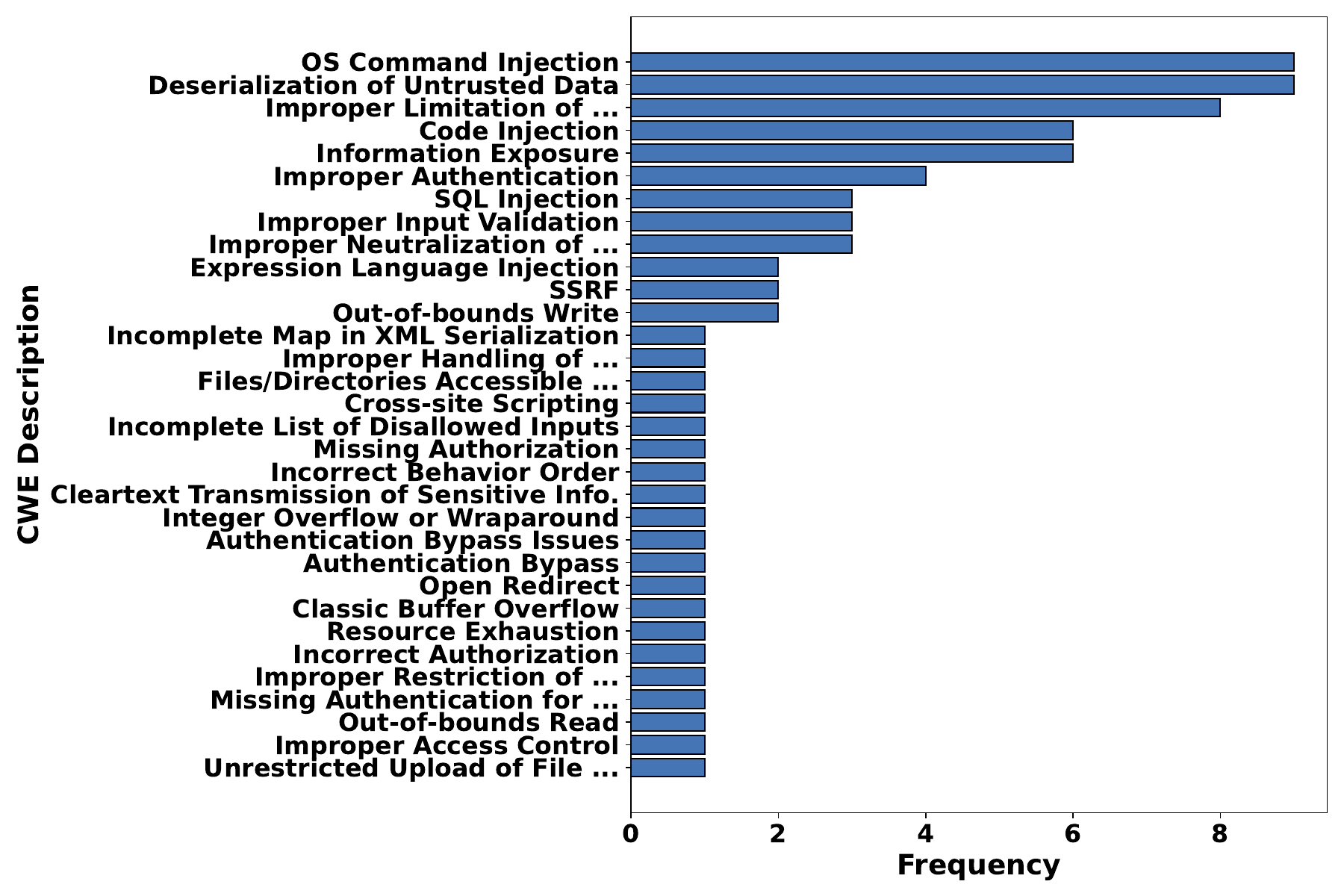}
    \Description[Coverage of CWE]{Coverage of CWE}
    \caption{Coverage of CWE}
    \label{fig:cwe_dist}
    % \vspace{-0.5em}
\end{figure}

\subsection{HackTheBox Benchmark Construction}
In selecting HackTheBox (HTB) challenges for evaluating \SysName, we aimed to create a diverse and representative set of targets that reflect real-world penetration testing scenarios. Our selection focused on key aspects such as operating system diversity, vulnerability relevance, and difficulty levels to ensure a comprehensive assessment.

To evaluate \SysName’s adaptability across different environments, we included both Linux and Windows machines. This diversity reflects the variety of systems encountered in real-world engagements and ensures that \SysName is tested across different exploitation techniques. The chosen machines also incorporate well-known vulnerabilities spanning a decade, allowing us to assess \SysName’s ability to handle both historical and contemporary exploits. For instance, Blue features the EternalBlue vulnerability (CVE-2017-0144), a widely known Windows SMB exploit, while Legacy includes MS08-067 (CVE-2008-4250), another critical SMB-based attack.

We selected machines across easy, medium, and hard difficulty levels to analyze \SysName’s performance in different attack scenarios. Easier challenges, such as Lame and Optimum, test fundamental exploitation techniques, while medium and hard machines, such as Stratosphere (CVE-2017-5638) and Reel (CVE-2017-0199), require deeper reconnaissance, multi-step attacks, and more advanced reasoning. This progression ensures that \SysName is evaluated not only on basic automation tasks but also on its effectiveness in handling complex, real-world pentesting challenges.

This carefully selected set of HTB challenges, as shown in Table~\ref{tab:practicality_pentestagent}, allows for a thorough assessment of \SysName’s automation capabilities, performance across different vulnerability types, and effectiveness in progressively complex penetration testing scenarios.

\begin{table}[htbp]
\caption{\SysName's performance among HackTheBox CTF challenges}
\label{tab:practicality_pentestagent}
\centering
\begin{tabular}{lccc}
\toprule
\textbf{Machine} & \textbf{Difficulty} & \textbf{Completed Stage} \\ \midrule
Sau             & Easy        &        2/3 (I.G, V.A)       &        \\ 
Pilgrimage     & Easy        &         1/3 (V.A)       &        \\ 
Lame            & Easy            &   3/3         &           \\ 
Topology         & Easy            &  3/3          &         \\ 
PC                  & Easy           & 3/3          &         \\ 
Blue                  & Easy           & 3/3      &             \\ 
Shocker                  & Easy           & 2/3 (V.A., E)        &           \\ 
Optimum                  & Easy            & 3/3       &            \\ 
Legacy                  & Easy            & 3/3         &          \\ 
Stratosphere            & Medium            &  2/3 (V.A., E)    &             \\ 
Reel                  & Hard               & 2/3 (V.A, E.)        &          \\ 
\bottomrule
\vspace{-1em}
\end{tabular}
\end{table}

\begin{table}[htbp]
\caption{\PentestGPT's performance among HackTheBox CTF challenges}
\label{tab:practicality_pentestGPT}
\centering
\begin{tabular}{lccc}
\toprule
\textbf{Machine} & \textbf{Difficulty} & \textbf{Completed Stage} \\ \midrule
Sau             & Easy        &        2/3 (I.G, V.A)       &        \\ 
Pilgrimage     & Easy        &         1/3 (V.A)       &        \\ 
Lame            & Easy            &   2/3 (V.A., E)         &           \\ 
Topology         & Easy            &  2/3 (V.A., E)         &         \\ 
PC                  & Easy           & 0/3          &         \\ 
Blue                  & Easy           & 2/3 (V.A., E)      &             \\ 
Shocker                  & Easy           & 0/3         &           \\ 
Optimum                  & Easy            & 3/3       &            \\ 
Legacy                  & Easy            & 3/3         &          \\ 
Stratosphere            & Medium            &  0/3     &             \\ 
Reel                  & Hard               & 1/3 (I.G.)        &          \\ 
\bottomrule
\vspace{-1em}
\end{tabular}
\end{table}

\subsection{Benchmark Coverage}
Our evaluation was conducted using a benchmark dataset comprising known vulnerabilities, which raises questions about the practicality in real-world scenarios.
Firstly, it is important to recognize that known vulnerabilities pose significant risks. Many organizations and institutions struggle with timely patching practices, contributing to vulnerable and outdated components ranking 6th on the OWASP Top 10 Web Application Security Risks.~\cite{owasp_top_10}
Additionally, while our benchmark dataset features known vulnerabilities, we selected environments based on their Exploit Prediction Scoring System (EPSS) scores. 
These scores reflect the likelihood of a vulnerability being exploited in real-world scenarios. 
% As shown in Fig.\ref{fig:epss_dist}, most environments in the benchmark dataset have EPSS scores above 90. 
The dataset's mean EPSS score is 79.58, with a median of 97.19, indicating that the vulnerabilities represented are highly likely to exist and be exploitable in practical settings.
Moreover, finding open datasets containing zero-day or even one-day vulnerable environments remains challenging. By focusing on known vulnerabilities with high EPSS scores, our evaluation ensures that \SysName operates within a realistic and credible context, assessing its effectiveness in addressing vulnerabilities that pose genuine risks to cybersecurity.

\section{Additional Evaluation Results}
\label{sec:addn_results}

The detailed pentesting performance of both systems on HackTheBox challenges are available in Table~\ref{tab:practicality_pentestagent} and Table.~\ref{tab:practicality_pentestGPT}.

\begin{figure}[htbp]
% \vspace{-0.5em}
    \centering
    \includegraphics[width=1.0\linewidth]{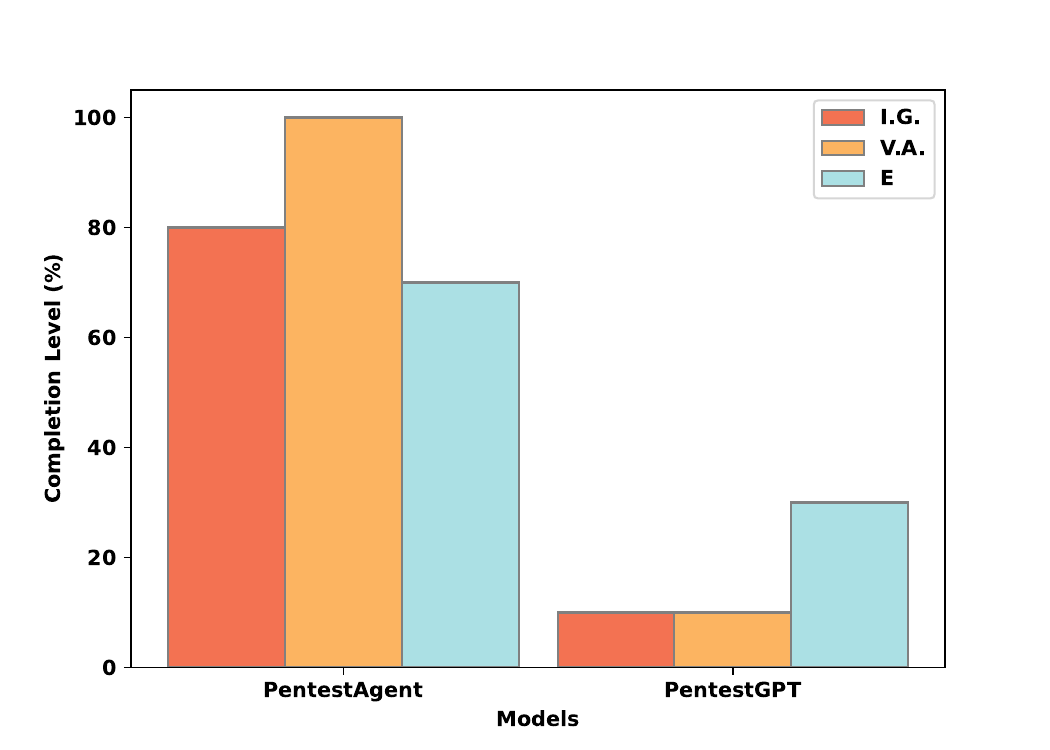}
    \caption{Comparison of completion levels on VulHub targets}
    \label{fig:completion_comparison_VulHub}
    % \vspace{-0.5em}
\end{figure}

\begin{figure}[htbp]
% \vspace{-0.5em}
    \centering
    \includegraphics[width=1.0\linewidth]{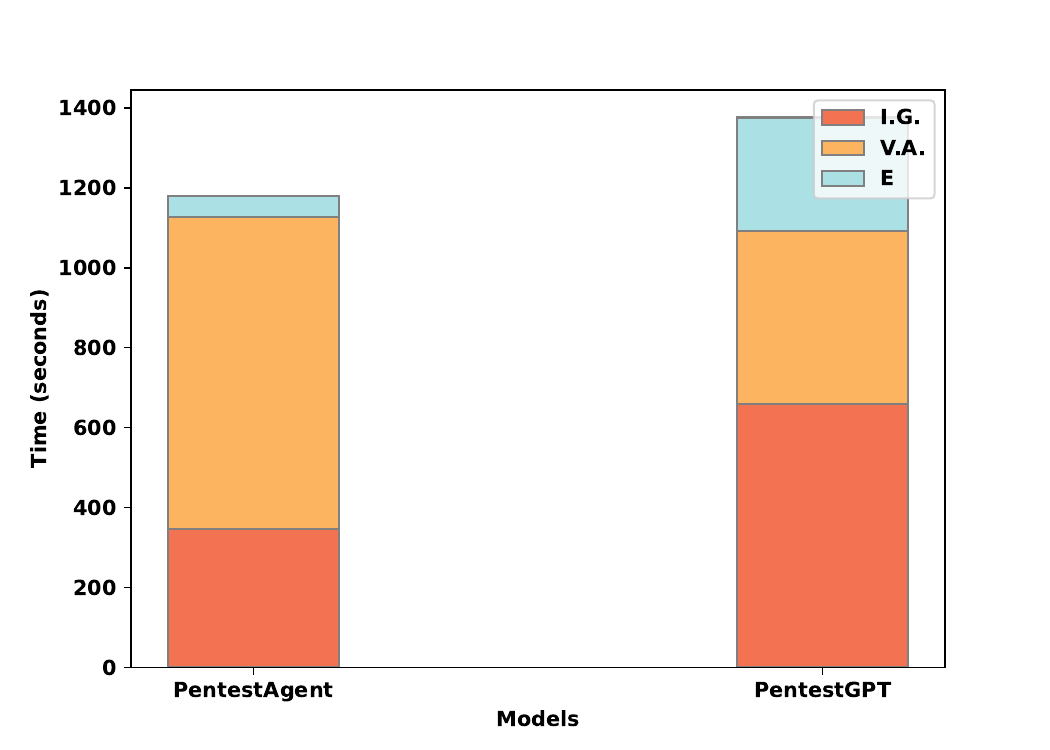}
    \caption{Comparison of average time spent on VulHub targets}
    \label{fig:overhead_comparison_vulhub}
    % \vspace{-0.5em}
\end{figure}

In addition to the HackTheBox evaluation, we compare \SysName and \PentestGPT on VulHub targets, analyzing both completion levels and overhead, as shown in Fig.~\ref{fig:completion_comparison_VulHub} and Fig.~\ref{fig:overhead_comparison_vulhub}. 

% \begin{figure*}[htbp]
% \vspace{-1em}
%      \centering
%      \begin{subfigure}[b]{0.48\textwidth}
%         \centering
%         \includegraphics[width=0.85\linewidth]{figures/completion_comparison_VulHub_cropped.pdf}
%         \caption{Comparison of completion levels on VulHub targets}
%         \label{fig:completion_comparison_VulHub}
%         \vspace{-1em}
%      \end{subfigure}
%      \hfill
%      \begin{subfigure}[b]{0.48\textwidth}
%         \centering
%         \includegraphics[width=0.85\linewidth]{figures/overhead_comparison_vulhub_cropped.pdf}
%         \caption{Comparison of average time spent on VulHub targets}
%         \label{fig:overhead_comparison_vulhub}
%         \vspace{-1em}
%      \end{subfigure}
%         \caption{Completion level and overhead comparison on VulHub targets.}
%         \label{fig:comparison_vulhub}
%         \vspace{-1em}
% \end{figure*}

% \begin{figure}[htbp]
% % \vspace{-0.5em}
%     \centering
%     \includegraphics[width=1.0\linewidth]{figures/completion_comparison_htb.pdf}
%     \caption{Comparison of completion levels on HackTheBox targets}
%     \label{fig:completion_comparison_htb}
%     % \vspace{-0.5em}
% \end{figure}

% \begin{figure}[htbp]
% % \vspace{-0.5em}
%     \centering
%     \includegraphics[width=1.0\linewidth]{figures/overhead_comparison_HTB.pdf}
%     \caption{Comparison of average time spent (seconds) on HackTheBox targets}
%     \label{fig:overhead_comparison_htb}
%     % \vspace{-0.5em}
% \end{figure}

\SysName significantly outperformed \PentestGPT across all penetration testing stages. In intelligence gathering, \SysName achieved an 80\% completion rate, compared to only 10\% for \PentestGPT, demonstrating its superior ability to extract relevant target information. In vulnerability analysis, \SysName completed 100\% of the tasks, whereas \PentestGPT again achieved just 10\%, highlighting its limited capability in identifying and assessing vulnerabilities. During exploitation, \SysName successfully completed 70\% of tasks, more than double \PentestGPT’s 30\%, confirming its stronger execution capabilities.

Efficiency is crucial in penetration testing automation, and \SysName consistently required less time than \PentestGPT in key stages. It completed intelligence gathering in 212.9 seconds, whereas \PentestGPT took 658.7 seconds, over three times longer. Similarly, in exploitation, \SysName finished in 58.6 seconds, compared to 283.5 seconds for \PentestGPT, demonstrating its more streamlined attack execution. While \SysName took longer in vulnerability analysis (698.8 seconds vs. 433.5 seconds), this additional time contributed to its higher success rate, ensuring a more accurate and actionable assessment.

The results confirm that \SysName is both more effective and more efficient than \PentestGPT. It achieves higher completion rates across all stages, particularly in intelligence gathering and vulnerability analysis, which are critical for successful exploitation. Moreover, its lower overhead in intelligence gathering and exploitation makes it a more scalable and practical solution for real-world penetration testing. While its vulnerability analysis takes slightly longer, this trade-off results in more reliable and successful attack execution, solidifying \SysName as a robust and efficient automated pentesting framework.

\end{document}